\documentclass{osa-article}
\usepackage{graphicx} 
\usepackage{caption}
\usepackage[skip=1pt]{caption}
\usepackage{subcaption}
\journal{oe} 
\articletype{Research Article}

\begin{document}
\title{Optical, Thermal, and Electrical Analysis of Perovskite Solar Cell with Grated CdS and Embedded Plasmonic Au Nanoparticles} 

\author{Ohidul Islam,\authormark{1}M. Hussayeen Khan Anik,\authormark{1} Sakib Mahmud,\authormark{1} Joyprokash Debnath,\authormark{1}, Ahsan Habib, \authormark{2} and Sharnali Islam\authormark{2,*}}

\address{\authormark{1}Department of Electrical and Electronic Engineering, Shahjalal University of Science and Technology, Sylhet 3114, Bangladesh\\
\authormark{2}Department of Electrical and Electronic Engineering, University of Dhaka, 1000, Dhaka, Bangladesh.\\
}

\email{\authormark{*}sharnali.eee@du.ac.bd} 


\begin{abstract}
 We propose a novel approach to enhance the performance of perovskite solar cells (PSCs) by incorporating grated Cadmium Sulfide (CdS) and plasmonic gold nanoparticles (Au NPs) into the absorber layer. The CdS grating acts as the electron transport layer and penetrates into the perovskite absorber layer, increasing the absorption of the active layer and reducing the electron-hole recombination rate. The plasmonic Au NPs enhance the absorption in the infrared region by scattering and trapping the incident light. We perform a coupled optical and electrical study that shows a significant improvement in the short circuit current density (J\textsubscript{SC}) and power conversion efficiency (PCE) of the PSC after introducing the CdS grating and plasmonic Au NPs. Specifically, we observe a 48\% increase in average optical absorption from 800 nm to 1400 nm and a 7.42 mA/cm\textsuperscript{2} increase in J\textsubscript{SC}. We also find that the PCE of the PSC is increased by 7.91\% when comparing the planar reference structure (without the CdS grating and the plasmonic Au NP). However, metal nanoparticles introduce ohmic losses and temperature rise in the solar cell. We analyze the non-radiative heat profile, electric field distribution, and temperature distribution across the PSC. We observe a temperature increase of approximately 14 K above the ambient temperature for the grated CdS layer with incorporated Au NPs, which is comparable to the temperature increase observed in the planar reference structure. Our results have the potential to pave the way for the development of highly efficient and stable PSCs in the future. 
\end{abstract}


\section{Introduction}
In recent years, there has been an increased focus on the development of cheap, renewable, and carbon-free energy sources due to the growing energy demand and the worsening impact of global warming \cite{lee2017review,nayak2019photovoltaic,siavash2020recent}. As a result, researchers have been exploring novel technologies to improve or possibly replace traditional silicon solar cells. Perovskite-based third-generation solar cells have gained significant interest due to their exceptional optical and electrical features, including long charge carrier diffusion length, high charge mobility, simple fabrication technique, and scalability \cite{burschka2013sequential,khattak2020numerical}. Halide-based perovskite solar cells (PSCs), such as CH\textsubscript{3}NH\textsubscript{3}PbX\textsubscript{3} (X = Cl, Br, and I), have shown remarkable photovoltaic performance, achieving electrical power conversion efficiency (PCE) above 22\% \cite{kumawat2016structural}. However, despite their impressive achievements, PSCs still face several challenges \cite{leong2016identifying}, such as the limited light absorption in the 700–1400 nm wavelength region \cite{sharma2022stability}, which results in recombination at the broken interface and inefficient use of incident light, leading to reduced efficiency. To achieve a PCE of 30\%, efficient light harvesting, and carrier transport are essential. Maximizing the short-circuit current density (J\textsubscript{SC}) by making full use of the sun's radiation and improving carriers' transit to enhance charge collection and reduce charge recombination are critical factors \cite{zhang2014electrochemical,chen2019recent}. Various methodologies and manufacturing techniques have been explored to increase the crystallinity, electronic characteristics of perovskite thin film, and charge transfer in PSCs, thereby improving their PCE and stability \cite{cao2020ordered,siavash2020recent}. \\

Efficient PSCs consist of a cathode/electron transport layer (ETL)/perovskite/hole transport layer (HTL)/anode. Perovskite optoelectronic devices contain several durable and stable metal oxide-based ETLs, including titanium dioxide (TiO\textsubscript{2}), tungsten oxide (WO\textsubscript{x}), and zinc oxide (ZnO), as well as cadmium sulfide (CdS) \cite{irandoost2020design}. One approach to improve PSCs is to incorporate a single-crystal grating array in the absorber layer and use cadmium sulfide (CdS) as the electron transport layer (ETL), which offers several advantages. First, the grating array provides a large surface area for exciton dissociation, which is the process by which an incoming photon generates a pair of mobile charge carriers (an electron and a hole) in the absorber layer. Second, the grating array creates a continuous channel for electron collection at the electrode, which improves the conductivity of the PSC. This results in a more efficient charge transfer and reduced charge recombination. Additionally, grated CdS arrays may help lower surface defects/traps in the PSC and modify energy levels to prevent the reverse recombination of electrons and holes. Finally, ordered grating array structures offer a direct carrier transport channel that facilitates charge transfer and reduces charge recombination. This is because the grating array creates a well-defined path for the charge carriers to travel, allowing them to reach the respective electrodes more efficiently  \cite{sun2018novel,rahman2022performance}.\\

The use of metallic nanoparticles (NPs) in photovoltaic devices has been extensively researched as a means of enhancing PSC's efficiency. Researchers have investigated various optical manipulation techniques such as diffraction grating, aperiodic dielectric stacking, and plasmonic metal nanoparticle implementation \cite{cui2016surface}. The incorporation of metallic NPs is believed to be one of the most efficient and practical methods of improving the ability of solar cells to absorb, scatter, and capture light \cite{brongersma2015plasmon,day2019improving}. Metallic NPs can trap resonant photons by inducing coherent oscillations of conduction band electrons at the surface, a phenomenon known as localized surface plasmon resonance (LSPR) \cite{irandoost2020design}. Plasmonic nanoparticles can also improve the photocurrents generated by PSCs and OSCs through LSPR \cite{alkhalayfeh2021enhancing}. The resonances in surface plasmons have large local-field amplitudes and strong far-field scattering, both of which are essential properties. The phenomenon of total internal reflection can exploit strong far-field scattering to increase the light absorption of thin-film solar cells. Furthermore, the high local fields in the region of plasmonic nanoparticles overlap with not only the metal but also the surrounding absorbing materials, resulting in an increase in absorption \cite{perrakis2019efficient}. The presence of plasmonic NPs in the absorber layer has been shown to increase parasitic absorption, which can serve as a recombination trap. However, several reports have indicated that the benefits of plasmonic NPs in enhancing photovoltaic performance outweigh the negative effects of parasitic absorption \cite{roopak2017light,pathak2016study}. Furthermore, the combination of plasmonic NPs and ETL gratings has not been extensively studied. To date, most studies have focused on either incorporating plasmonic NPs or nanostructures into the ETL and the absorber layer. However, the potential benefits of combining these two approaches warrant further investigation.\\

In this work, we use a simulation study to improve the performance of perovskite solar cells by combining two realistic ways: we replace the planar electron transport layer with a CdS grating and add gold nanoparticles (NPs) in the spaces between the gratings. We numerically develop a planar PSC composed of ITO/CdS/MAPbI\textsubscript{3}/Spiro-OMeTAD/Au, where CdS, Spiro-OMeTAD, MAPbI\textsubscript{3}, and Au serve as the ETL, HTL, perovskite absorber layer, and back metal contact, respectively. In the second stage, we add a CdS grating to the absorber layer to improve the device's performance and lower the likelihood of electron-hole recombination. We insert plasmonic Au NPs between CdS gratings to enhance light absorption beyond the visible spectrum. We investigate the optical absorption of the PSCs and examine the electrical properties such as PCE, fill factor (FF), and J\textsubscript{sc}. After incorporating Au NPs into our PSC, we observe over 48\% increased optical absorption in the infrared band of the sun's light. We achieve a significant improvement in photocurrent and PCE by 24.2\% and 56.6\%, respectively, where J\textsubscript{sc} increases from 30.67 mA/cm\textsuperscript{2} to 38.09 mA/cm\textsuperscript{2} and PCE increases from 13.96\% to 21.87\% when comparing the initial structure with the final structure in the electrical simulation. We perform a thorough investigation to learn the impact of CdS grating width, gold nanoparticle size, and location of nanoparticles. We use the coupled optical and thermal simulation to look at the spreading of heat and temperature. Our analysis shows that the temperature rise for all structures is approximately 14 K from the ambient temperature of 300 K, highlighting the applicability of our proposed PSC with grated CdS and embedded plasmonic nanoparticles. We also find that the grated CdS with Au NPs exhibits a notable improvement in quantum efficiency(QE) over a wide range of wavelengths, from 300 nm to 1200 nm, compared to the planar structure. We discuss a potential fabrication scheme in our manuscript. Our work paves the way for advancements in the field of optical and electrical performance enhancement for future solar cells due to its effective absorption of light outside the visible spectrum, increased photocurrent and PCE, and meticulous analysis of heat and temperature distribution across the PSC.


\section{Numerical Method and Device Structure}
\subsection{Methodology}
The aim of this study is to investigate the impact of CdS grating and Au NPs on the performance of a perovskite solar cell (PSC) using optoelectronic simulation methods. Two simulation tools were employed, namely Finite Difference Time Domain (FDTD)-based Ansys Lumerical software and SCAPS-1D. The FDTD method was used to solve Maxwell's equations at each mesh node, which helped to determine the optical absorption and photogeneration rate. Three fundamental semiconductor equations, including Poisson's equation, the drift-diffusion equation, and the carrier continuity equation, were solved to calculate important parameters such as the current density - voltage (I-V) characteristic and the spectral response of the solar cell. The interaction between plasmonic nanoparticles and the incident electromagnetic (EM) field was found to be critical to the device's performance. The electric field distribution was calculated using the FDTD method, and AM1.5G spectra were used as the input light power. The simulations were carried out in the wavelength range between 300 nm and 1400 nm, with a bandgap of around 1.5 eV for the absorber layer. This means that light with a wavelength longer than 800 nm cannot be absorbed by the absorber layer, and the plasmonic NPs were used to increase the amount of light absorbed outside of the visible spectrum. Refractive index (n) and extinction coefficient (k) values for materials such as CdS, Spiro-OMeTAD, and MAPbI\textsubscript{3} were obtained from previously published research \cite{chen2015optical,rahman2022performance,baum2013determination,palik1998handbook}. For Au metal NPs, the Palik data was used. To create a 3D model of the suggested PSC, perfectly matched layer (PML) boundary conditions were used for the Y and Z directions of the solar cell, while periodic boundary conditions were used for the X direction. The position-dependent absorptance A(r,$\omega$) was computed using the divergence of the Poynting vector normalized over the incoming radiation power, which was provided in the study by Heidarzadeh et al. \cite{heidarzadeh2021design}

\begin{equation}
    \textbf{A}(\textbf{r}, \omega)=\frac{1}{2} \frac{\operatorname{real}(\overrightarrow{\nabla \cdot \textbf{P})}}{\textbf{P}_{i n}}=\frac{1}{2} \frac{\omega \varepsilon_0 \operatorname{\textbf{I}m}(\varepsilon(\textbf{r}, \omega))|\textbf{E}(\textbf{r}, \omega)|^2}{\textbf{P}_{i n}}
    \label{eqn:A}
\end{equation}

where  $\epsilon$ is the dielectric constant, $\omega$ is the angular frequency, and E is the electric field vector. The absorbance in MAPbI\textsubscript{3}, however, may be expressed as follows\cite{heidarzadeh2021design}

\begin{equation}
    \textbf{A}_{\text {total }}(\lambda)=\omega \varepsilon_0 \int_V|\textbf{E}(x, y, z, \lambda)|^2 \textbf{n}(\lambda) \textbf{k}(\lambda) d \textbf{V}
    \label{eqn:A_total}
\end{equation}

where $\epsilon_0$, $\lambda$, n, k, and V represent the vacuum permittivity, vacuum wavelength, real and imaginary components of the refractive index, and the integral volume computed inside it, respectively. Nevertheless, the absorption spectra of metal nanoparticles are examined independently by restricting the range of integrals within the NPs. Using the Poynting vector divergence, it is possible to calculate the value of absorption power per unit volume as follows\cite{rahman2022performance}

\begin{equation}
    \textbf{P}_{\text{abs}}=-\frac{1}{2} \omega\left|\overrightarrow{\textbf{E}_{\text{op}}}(\vec{\textbf{r}}, \omega)\right|^2 \textbf{I}\{(\vec{\textbf{r}}, \omega)\}
    \label{eqn:P_abs}
\end{equation}

where $\overrightarrow{\textbf{E}_{\text{op}}}$ denotes electric field orientation and $\textbf{I}\{(\vec{\textbf{r}}, \omega)\}$ represents the imaginary part of the complex permittivity of each material. Next, using the acquired P\textsubscript{abs} and the following Equations\cite{rahman2022performance}, we can determine the rate of electron-hole pair generation $\textbf{G}(\vec{r})$ inside the various layers.

\begin{equation}
    \textbf{G}(\vec{r})=\int g(\vec{r}, \omega) d \omega
    \label{eqn:G_r}
\end{equation}

\begin{equation}
    g(\vec{r}, \omega)=\frac{\textbf{P}_{\text{abs}}}{\hbar \omega}=-\frac{\pi}{h}\left|\overrightarrow{\textbf{E}_{\text{op}}}(\vec{\textbf{r}}, \omega)\right|^2 \textbf{I}\{\epsilon(\vec{\textbf{r}}, \omega)\}
    \label{eqn:g_r_w}
\end{equation}

where, h denotes the Planck's constant. After completing the optical simulation in Lumerical FDTD, we exported the resultant absorption data to SCAPS-1D for the electrical simulation. Using the fundamental photovoltaic equations for each layer, which are controlled by user-defined architectural and material factors, SCAPS is able to simulate solar responses. Three fundamental equations regulate the behavior of carriers in solar devices: the Poisson equation, the carrier continuity equation, and the drift-diffusion charge transport model equations. The Poisson equation describes the connection between the electric field produced at a P-N junction and the space charge density, and it may be represented as\cite{anik2022comparative}  

\begin{equation}
    \frac{\partial^2 \psi(x)}{\partial x^2}=\frac{q}{\epsilon_0 \epsilon_r}\left[p(x)-n(x)+N_D^{+}-N_A^{-}+p_{\text {trap }}(x)-n_{\text {trap }}(x)\right]
    \label{eqn:del_si}
\end{equation}

In this case, $\psi (x)$ is the unidirectional electrostatic potential, $q$ is the electrostatic charge, and r is the relative permittivity of the semiconductor material. Hole and electron densities are denoted by $p(x)$ and $n(x)$, whereas trapped densities are shown by $p_{trap} (x)$ and $n_{trap} (x)$, respectively. Also, the following are the electron and hole carrier continuity equations\cite{anik2022comparative}.

\begin{equation}
    -\frac{1}{q} \frac{\partial J_n}{\partial x}=R_n(x)-G(x)
    \label{eqn:current_density_n}
\end{equation}

\begin{equation}
    \frac{1}{q} \frac{\partial J_p}{\partial x}=R_p(x)-G(x)
    \label{eqn:current_density_p}
\end{equation}

Here, the current density of electrons and holes is denoted respectively by $J_{n}$ and $J_{p}$. The rate of photogeneration of electron-hole pairs (EHPs) is denoted by the symbol G. The electron and hole recombination rates, respectively, are denoted by $R_n$ and $R_p$.

In the carrier transport model, which is used to determine the electron and hole densities, the total current density J is found to be equal to the combination of the drift and diffusion current densities\cite{moiz2021design}. 

\begin{equation}
    J=J_n+J_p 
    \label{eqn:total_J}
\end{equation}
\begin{equation}
    J_n=q n(x) \mu_n \frac{d \psi(x)}{d x}+q D_n \frac{d n(x)}{d x}
    \label{eqn:J_n}
\end{equation} 
\begin{equation}
    J_p=q n(x) \mu_p \frac{d \psi(x)}{d x}+q D_p \frac{d n(x)}{d x}
    \label{eqn:J_p}
\end{equation}

The diffusion coefficients for electrons and holes are denoted by the symbols $D_n$ and $D_p$, whereas the mobility of electrons and holes is expressed by the symbols $\mu_n$ and $\mu_p$, respectively. The incident monochromatic photon's frequency is set at $10^{18} s^{-1}$. In SCAPS simulation, the shunt and series resistances of solar cells are fixed to 5000 Ohm$.$cm\textsuperscript{2} and 1 Ohm$.$cm\textsuperscript{2} accordingly. This study utilizes conventional AM 1.5G lighting at an  ambient temperature (T) of 300 K. 

In addition to optical and electrical systems, photovoltaic devices are also thermodynamic systems; these three physical domains are strongly connected. This study also investigates the impact of optical absorption-induced heating on the electrical performance of the solar cell in the normal mode of operation. To get an understanding of the non-radiative heat and temperature distribution in the cell structure, we ran thermal simulations with Lumerical HEAT by linking the thermal module to the optical module and adding a heat source in Lumerical HEAT based on the optical absorption reported in Lumerical FDTD. The temperature distribution is determined by the solution to the differential equation of steady-state conduction with internal heat generation\cite{zandi2020numerical}: 

\begin{equation}
    -k \nabla^2 T+Q=\rho_p C_p \frac{d T}{d t}
    \label{eqn:k_nabla_T}
\end{equation}

where k is a material's temperature-dependent thermal conductivity (W/(mK)), Cp is its specific heat (J/(kg-K)), and $\rho_d$ is its density. Q determines the amount of net energy absorbed at the sample surface exposed to concentrated solar radiation. It is also known as the source term. The amount of heat produced by the Joule effect is determined only by the product of the electric intensity and the current density.

\subsection{Device Structure}
\begin{figure}[!ht]
    \centering
    \includegraphics[width=12cm]{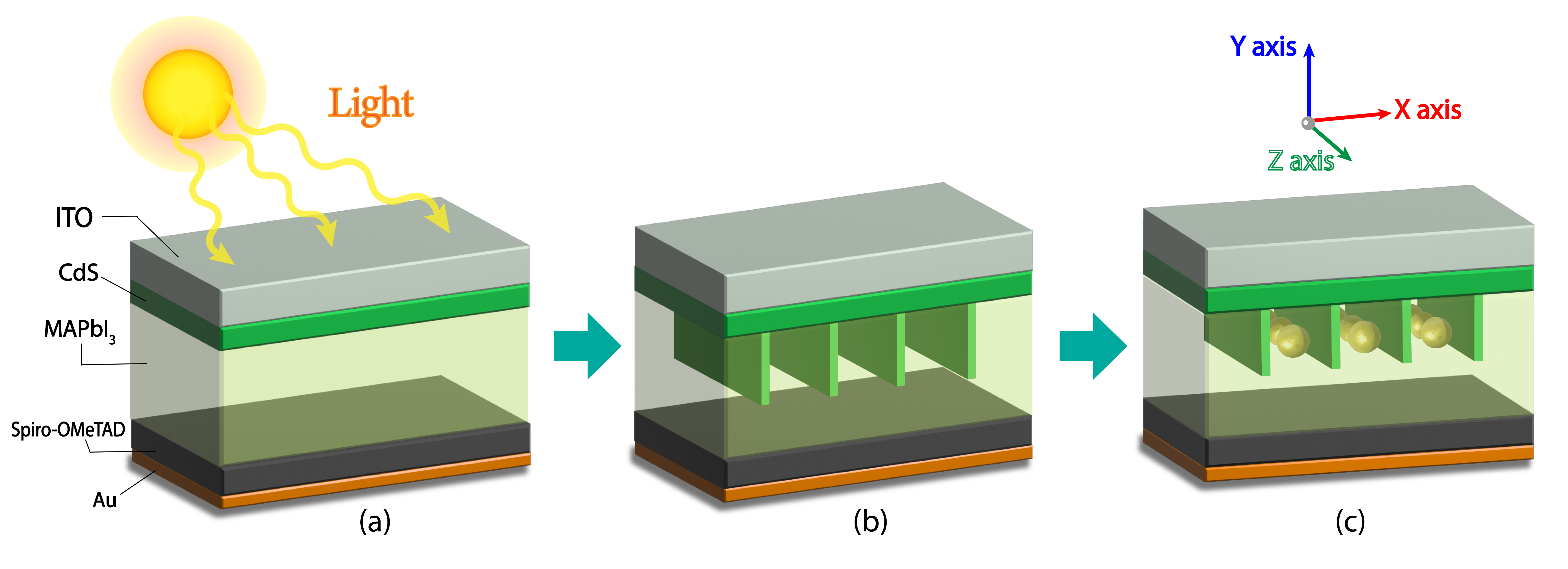}
    \caption{(a) The initial planar structure of the PSC, which serves as the reference design. (b) The PSC's development is shown by the introduction of a CdS grating into the absorber layer. (c) The final design of the PSC is shown by incorporating plasmonic spherical Au nanoparticles placed between the grated CdS, further enhancing its performance. These 3D illustrations provide a visual representation of the evolution of the PSC's structure and material composition, allowing for a better understanding of the design and fabrication process.}
    \label{fig:struc_development}
\end{figure}

We provide a detailed description of the materials and geometries used in our study and present comprehensive 3D depictions of the device structures. Figure \ref{fig:struc_development}(a) shows the planar structure, while Fig. \ref{fig:struc_development}(b) illustrates the incorporation of a CdS grating. Finally, Fig. \ref{fig:struc_development}(c) displays the structure consisting of both grating and plasmonic Au nanoparticles. These figures provide an overview of the complex geometries and material compositions used in our study, allowing for a better understanding of the design and fabrication process.

The planar structure used in this study implements an n-i-p configuration with a CdS electron transport layer (ETL), a MAPbI\textsubscript{3} absorber layer, and a Spiro-OMeTAD hole transport layer (HTL), with tin-doped indium oxide (ITO) used as the cathode layer and gold (Au) as the back contact. ITO was chosen as the cathode layer due to its high optical transparency, chemical inertness, excellent substrate adherence, and high conductivity \cite{yu2016indium}. Gold was selected as the back contact due to its superior electrical conductivity and compatibility with perovskite materials \cite{behrouznejad2016study}. These choices of materials and layer orientations are critical to achieving the desired performance and efficiency in the PSC and were carefully considered based on their known properties and prior research in the field.

\begin{figure*}[ht!]
    \centering
        \begin{subfigure}[b]{0.38\textwidth}
        \includegraphics[width=\textwidth]{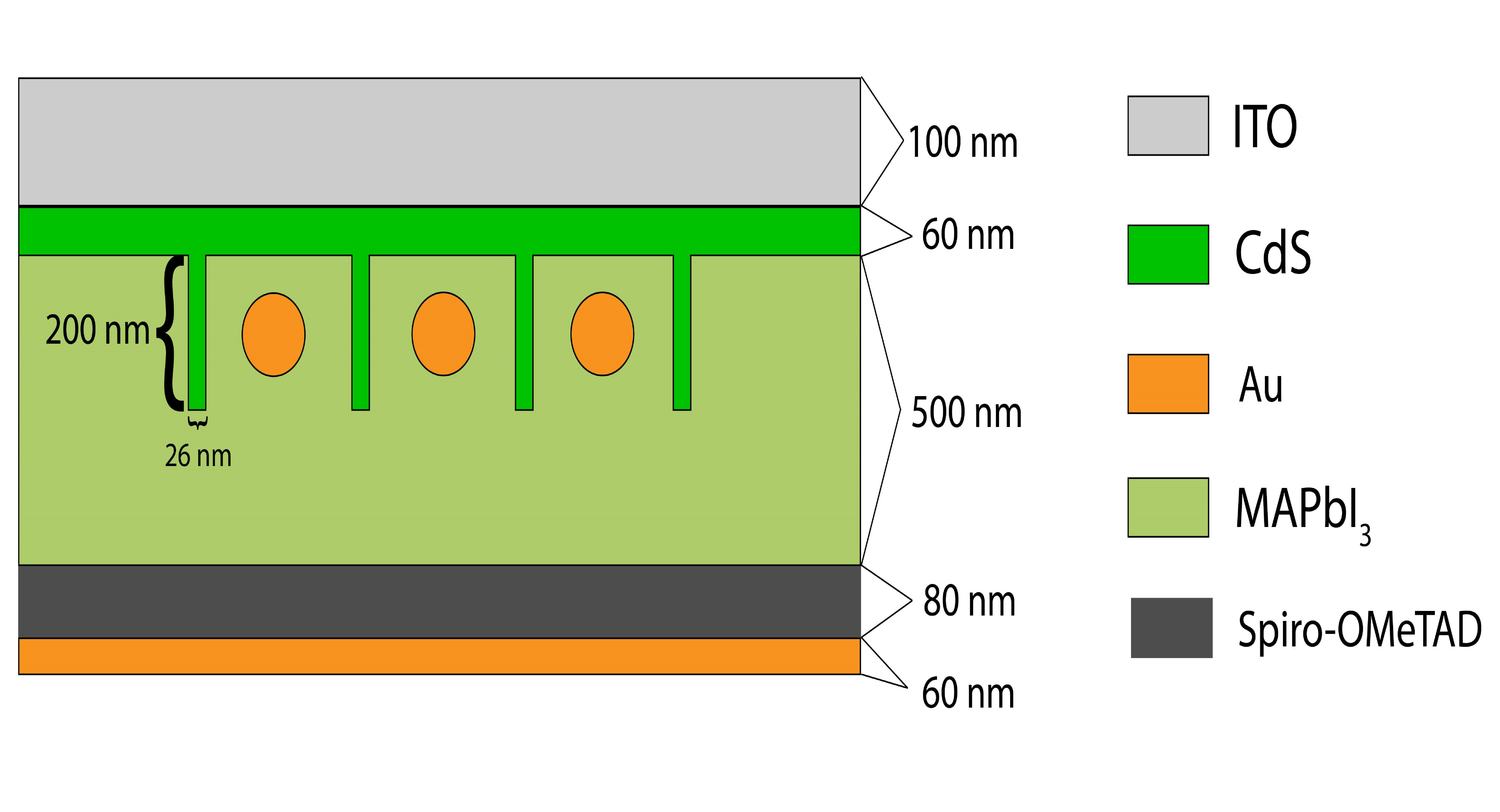}
        \caption{}
       \label{fig:2d_struc}
        \end{subfigure}
        \begin{subfigure}[b]{0.38\textwidth}
        \includegraphics[width=\textwidth]{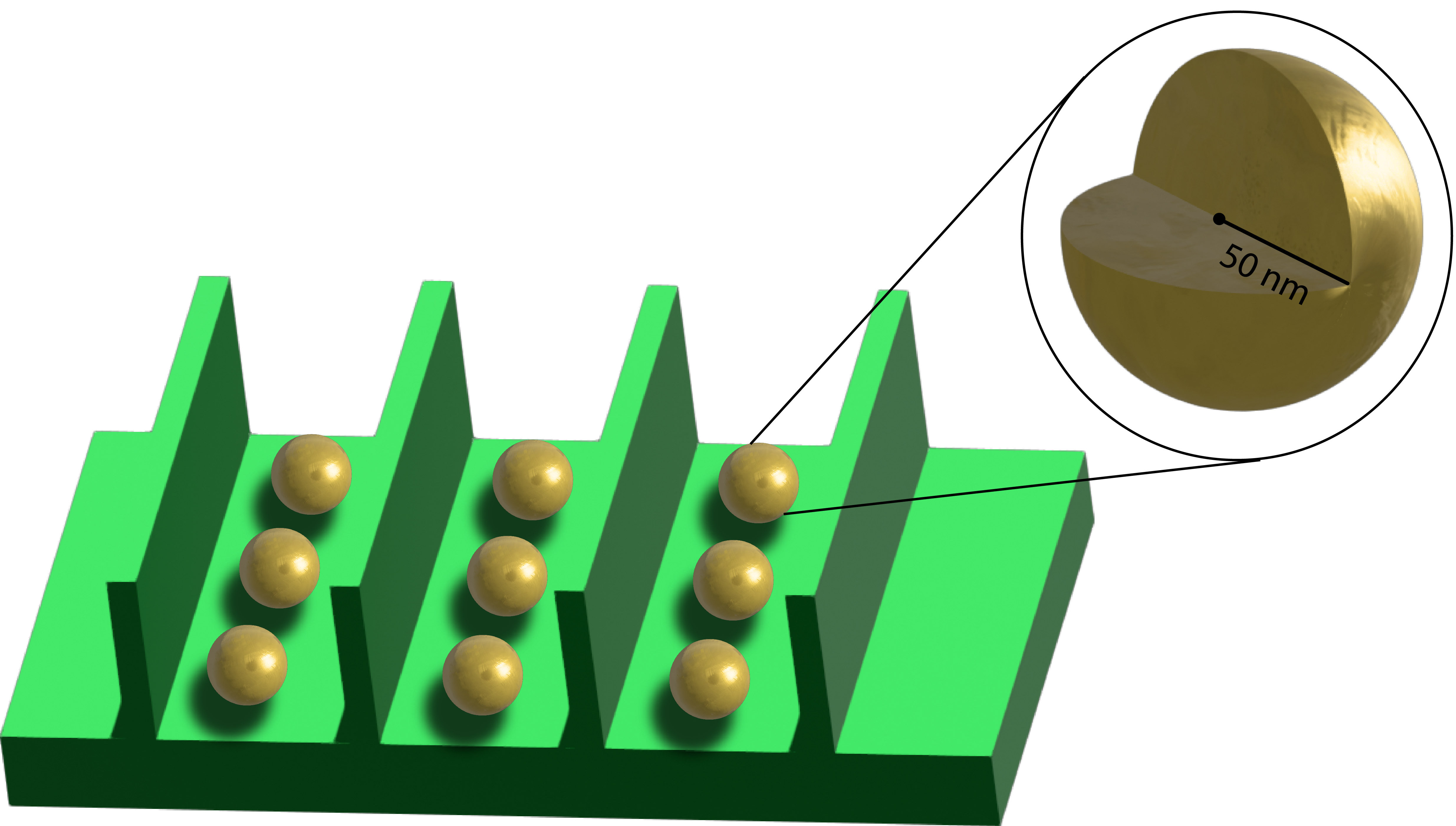}
        \caption{}
        \label{fig:backside}
        \end{subfigure}
    
    \caption{(a) A 2D view of the final structure, with detailed dimensional specifications provided. This figure provides a detailed view of the final design, allowing for a better understanding of the overall geometry and material composition of the PSC. (b) A rotated bottom view of the grated CdS electron transport layer (ETL) with Au nanoparticles (NPs) structure. This view provides a closer look at the grated CdS ETL with the incorporated Au NPs, which play a critical role in enhancing the PSC's performance. Together, these 2D depictions offer valuable insights into the device's structure and material composition, aiding in the interpretation of the experimental results.}
    \label{fig:2d_backside}
\end{figure*}

The dimensions of each CdS grating are specified in a 2D view presented in Fig. \ref{fig:2d_struc}, and a more detailed illustration of the orientation of the CdS gratings and Au NPs is given in Fig. \ref{fig:backside}. Spherical Au NPs with a radius of 50 nm are introduced between the CdS gratings in the absorber layer, resulting in further improvements in light absorption.

Table \ref{tab:simulation_prop} provides important simulation parameters for all the corresponding layers, including thickness, bandgap, electron affinity, relative dielectric permittivity, the carrier density of states, and other features necessary for modeling a solar cell device. The references for the corresponding parameters are also cited at the bottom of the table.

\renewcommand{\arraystretch}{1.1}
\begin{table}[!ht]
    \small
    \centering
    \caption{Input Electrical Parameters for the proposed PSC}
    \label{tab:simulation_prop}
    
    \begin{tabular}{>{\centering\arraybackslash}m{5.1cm}  >{\centering\arraybackslash}m{1.3cm} >{\centering\arraybackslash}m{1.3cm} >{\centering\arraybackslash}m{4cm} >{\centering\arraybackslash}m{1.3cm}}
        \hline
        \textbf{Parameters} & \textbf{ITO} & \textbf{CdS}	& \textbf{MAPbI\textsubscript{3}} & \textbf{Spiro-OMeTAD} \\
        \hline
        Thickness (nm) & 100 &	60 & 500 & 80 \\
        
        Bandgap (eV) & 3.6 & 2.42 & 1.50 & 3.06 \\
        
        Electron Affinity (eV) & 4.5 & 3.75 & 3.93 & 2.1 \\
        
        Permittivity & 8.9 & 10 & 10 & 3\\
        
        Effective Density of States at CB (cm\textsuperscript{-3}) & 2.2×10\textsuperscript{18} & 1.17×10\textsuperscript{19} & 2.75×10\textsuperscript{18} & 2.8×10\textsuperscript{19} \\
        
        Effective Density of States at VB (cm\textsuperscript{-3}) & 1.8×10\textsuperscript{19} & 4.12×10\textsuperscript{18} & 3.9×10\textsuperscript{18} & 1×10\textsuperscript{19} \\
        
        Electron Mobility (cm\textsuperscript{-2}/Vs) & 10 & 160 & 1.0 & 1×10\textsuperscript{-4} \\
        
        Hole Mobility (cm\textsuperscript{-2}/Vs) & 10 & 15 & 1.0 & 1×10\textsuperscript{-4} \\
        
        Donor Doping Density, N\textsubscript{D} (cm\textsuperscript{-3}) & 1×10\textsuperscript{19} & 1×10\textsuperscript{18} & 1×10\textsuperscript{9 }& | \\
        
        Acceptor Doping Density, N\textsubscript{A} (cm\textsuperscript{-3}) & | & | & 1×10\textsuperscript{9} & 4×10\textsuperscript{18} \\
        
        Defect Density, N\textsubscript{t} (cm\textsuperscript{-3}) & 1×10\textsuperscript{11} & 1×10\textsuperscript{11} & 1×10\textsuperscript{11} & 1×10\textsuperscript{11} \\
        
        Electron Thermal Velocity (cm/s) & 1×10\textsuperscript{7} & 1×10\textsuperscript{7} & 1×10\textsuperscript{7} & 1×10\textsuperscript{7} \\
        
        Hole Thermal Velocity (cm/s) & 1×10\textsuperscript{7} & 1×10\textsuperscript{7} & 1×10\textsuperscript{7} & 1×10\textsuperscript{7} \\
        
        Capture Cross Section Electron (cm\textsuperscript{2}) & 1×10\textsuperscript{-16} & 1×10\textsuperscript{-15} & 1×10\textsuperscript{-15} & 1×10\textsuperscript{-14} \\
        
        Capture Cross Section Holes (cm\textsuperscript{2}) & 1×10\textsuperscript{-16} & 1×10\textsuperscript{-15} & 1×10\textsuperscript{-15} & 1×10\textsuperscript{-14} \\
        
        Radiative recombination coeff. (cm³/s) & | & 1.02×10\textsuperscript{-10} & 1.1×10\textsuperscript{-7} (Planar structure) 1.1×10\textsuperscript{-9} (Grated CdS)
        1.1×10\textsuperscript{-8} (Grated CdS \& NPs) & 1.0×10\textsuperscript{-10} \\
        
        Reference & \cite{azri2019electron} & \cite{rahman2022performance} & \cite{jamal2019effect} &  \cite{he2021design} \\
        \hline
    \end{tabular}

\end{table}

\subsection{Fabrication Possibility and Discussion}
To ensure the practical implementation of the solar cell device, particularly incorporating gratings and nanoparticles, careful consideration must be given to a suitable and feasible fabrication process. Four design aspects must be taken into account while modeling the optimum design for dielectric grating and plasmonic nanoparticles. Firstly, the selected grating design should be simple and universally manufacturable on a large scale at an economical cost. Secondly, to prevent an increase in the width of the solar cells and the total cost, the height of the grating formation should be smaller than the breadth of the bottom active layer. Thirdly, to achieve Mie resonance with a high scattering cross-section, the size of the dielectric grating must be subwavelength. Finally, the grating and nanoparticle diameters must also be subwavelength to allow for the localization of surface plasmon resonance (LSPR) to guide and confine light. These considerations are crucial for ensuring the practicality and scalability of the solar cell device's fabrication process while maintaining its enhanced light absorption properties \cite{abdelraouf2018front}.

\begin{figure}[!ht]
    \centering
    \includegraphics[width=12cm]{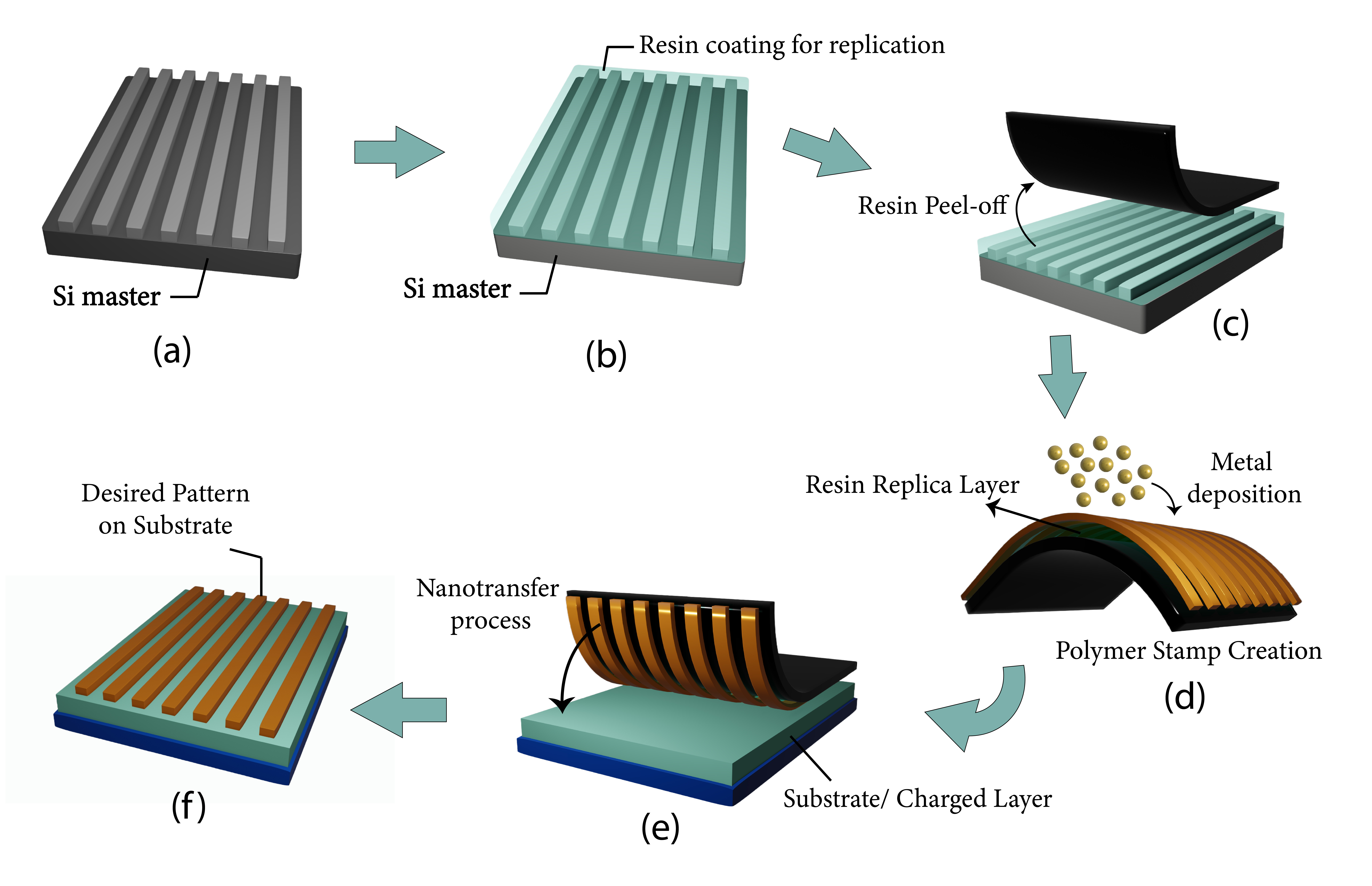}
    \captionsetup{justification=raggedright,singlelinecheck=false}
    \caption{Schematic representation of the nanotransfer lithography (NTL) fabrication process. In step (a), a Si master is shown, which is replicated by applying a polymer resin coating in step (b). In step (c), the resin coating is peeled-off. The preparation of the polymer stamp is shown in step (d) where the polymer resin material must be durable enough to withstand the mechanical and thermal impact of the transfer and deposition process. Desirable plasmonic metal (Au, Ag, Al) or dielectric (CdS, SiO\textsubscript{2}) is deposited over the polymer stamp, which will then be transferred onto the desired substrate or material layer in step (e). Finally, in step (f), the outcome after the material deposited on the flexible polymer stamp has been transferred onto the desired layer is shown.}
    \label{fig:fabrication}
\end{figure}

Various fabrication techniques are available today for the efficient implementation of devices incorporating gratings and nanoparticles. Nanotransfer lithography (NTL), chemical vapor deposition (CVD), and chemical bath deposition (CBD) are some of the commonly used methods. NTL has been successful in constructing nanostructures for photon management in optoelectronic devices due to its low cost, quick turnaround time, flexibility, and reproducibility. However, the fragility of typical Si masters used in this process poses a significant limitation. Polydimethylsiloxane (PDMS) masters, which are often used as Si alternatives, have a complicated fabrication process. The NTL fabrication process involves the replication of the Si master by applying a polymer resin coating, followed by the preparation of the polymer stamp, as illustrated in Fig. \ref{fig:fabrication}. 
\\
The desirable plasmonic metal (Au, Ag, Al) or dielectric (CdS, SiO\textsubscript{2}) is deposited over the polymer stamp, which is then transferred onto the desired substrate or material layer. Our work solely focuses on a simulation-based approach to obtain results, but based on the current literature, it is feasible to effectively fabricate the proposed design using NTL due to its cost-effectiveness, flexibility, and controllability. For Au-NPs synthesis, a kinetically controlled seeded growth strategy can be implemented, which executes the chemical reduction of HAuCl\textsubscript{4} by sodium citrate. The incorporation of processed NPs into a perovskite layer can be achieved via methods such as spin coating, electrodeposition, etc.

\section{Result and Discussion}
\subsection{Structure modification (from planar to grated CdS with Au NP)}

\begin{figure*}[ht!]
    \centering
        \begin{subfigure}[b]{0.39\textwidth}
        \includegraphics[width=\textwidth]{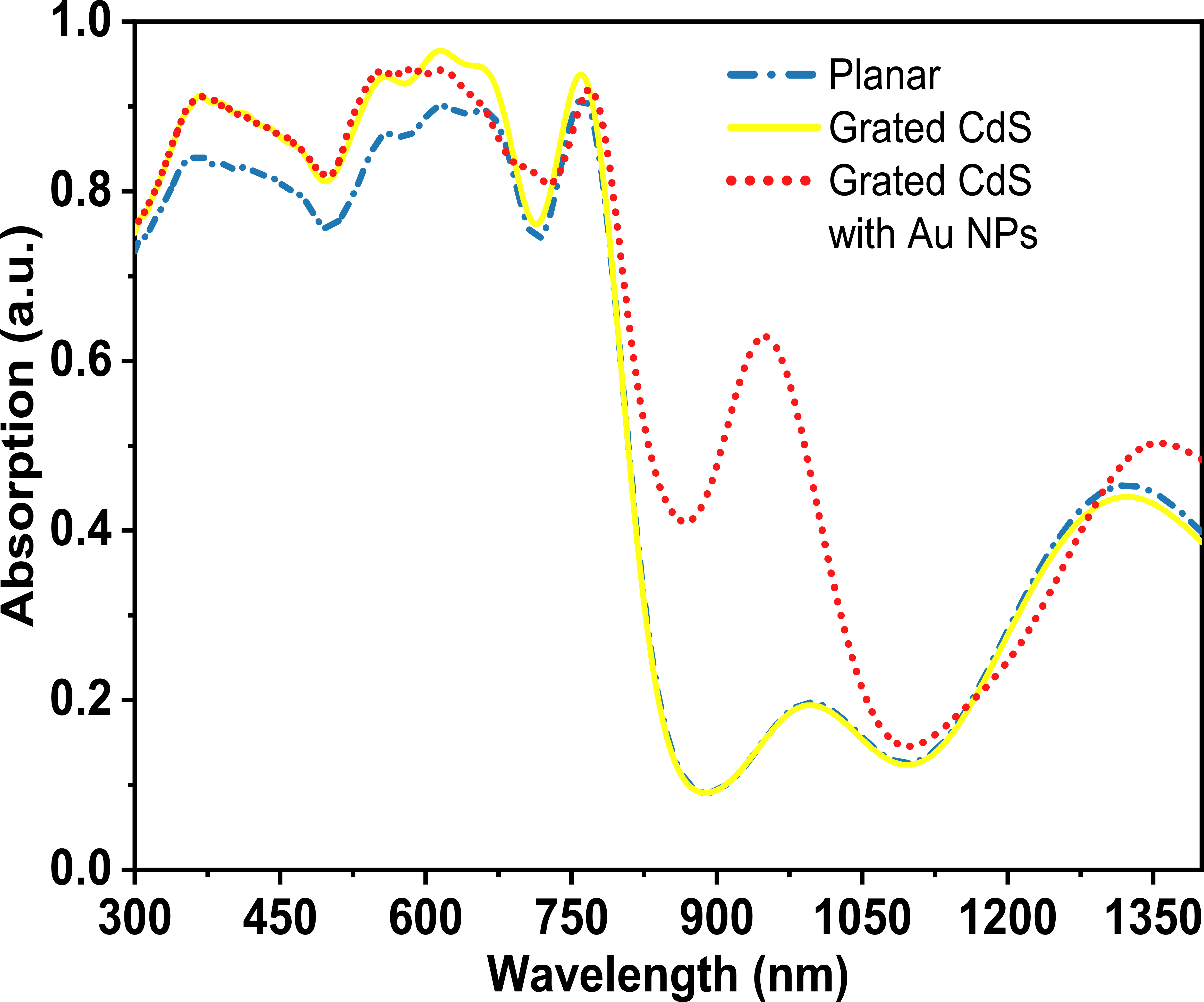}
        \caption{}
       \label{fig:ABS_main}
        \end{subfigure}
        \begin{subfigure}[b]{0.4\textwidth}
        \includegraphics[width=\textwidth]{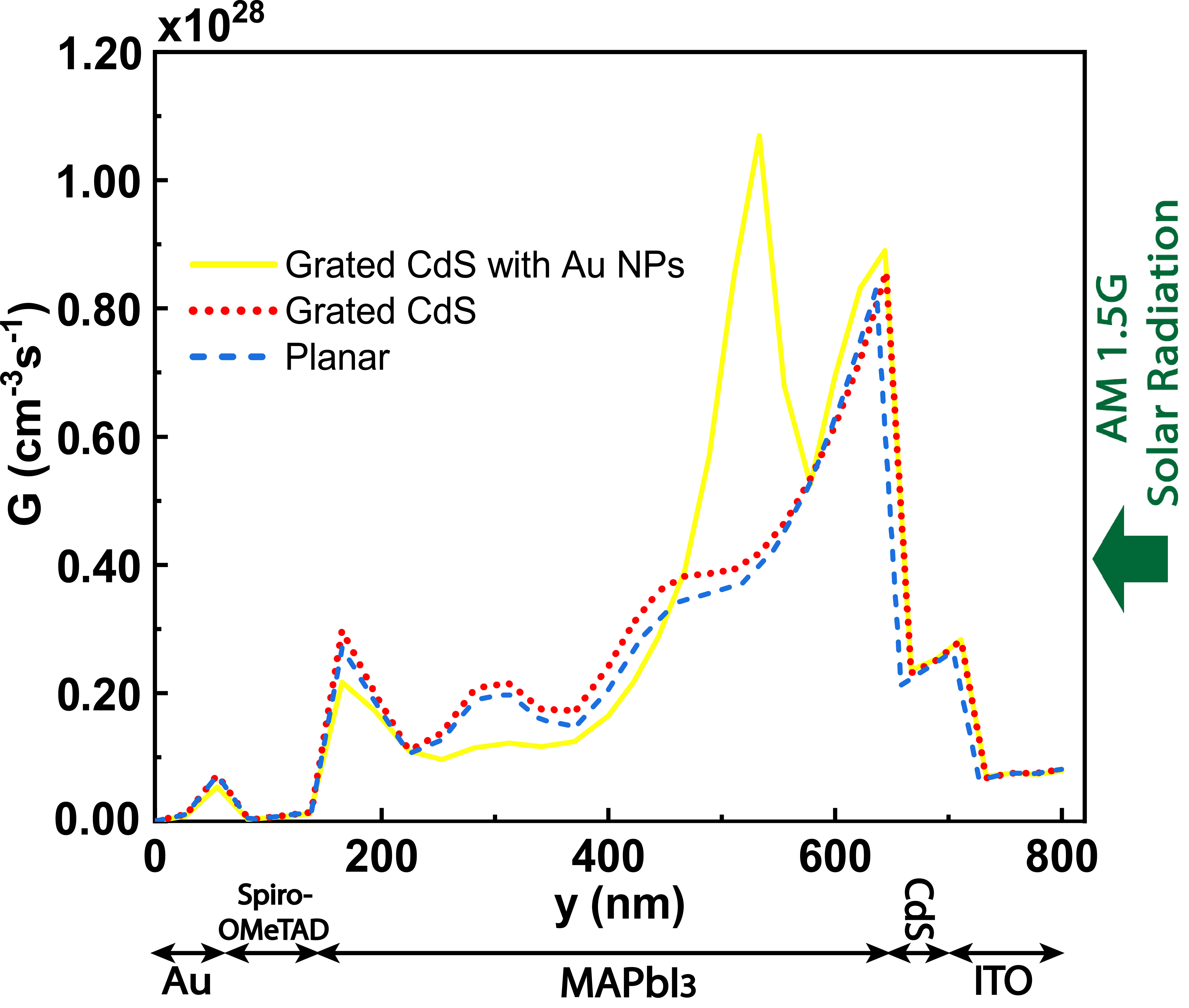}
        \caption{}
        \label{fig:gen_main}
        \end{subfigure}
    
    \caption{Comparison of the optical absorption spectrum and generation rate curves for three different structures of perovskite solar cells (PSCs), ranging from a planar structure to a structure with grated CdS and gold nanoparticles (Au NPs). (a) The optical absorption spectrum shows the amount of light absorbed by each structure across a range of wavelengths. (b) The generation rate curves represent the rate at which electron-hole pairs are generated within each structure. }
    \label{fig:ABS_gen_main}
\end{figure*}

As light beams travel from the ITO to the back contact of Au, they undergo absorption in various parts of the structure. However, it is crucial to direct the generated electron-hole pair into the CdS and Spiro-OMeTAD regions, respectively, to produce a photocurrent. To enhance this optoelectronic effect, one approach is to utilize a CdS grating. Gratings provide multiple opportunities for the incoming light to bounce off between grating surfaces, which increases the optical path as well as the absorption probability. From  Fig. \ref{fig:ABS_main}, it is apparent that utilizing grated CdS increases the absorption to nearly 1.0 in the visible range of light, indicating that almost all incoming visible light hitting the ITO surface is absorbed by the structure.

Adding grated CdS does not improve the absorption of the PSC beyond 750 nm. However, the addition of gold nanoparticles between the grated CdS allows for further enhancement of absorption into the IR region. These nanoparticles scatter light with wavelengths close to their plasmon resonance, and plasmon stimulation can boost absorption and carrier generation. As shown in Fig. \ref{fig:ABS_main}, the absorption curve improves with the addition of Au NPs, as evidenced by the peak at 945 nm. This improvement is also evident in the generation rate curve shown in Fig. \ref{fig:gen_main}, where the generation rate of a solar cell is the ratio of the total number of electrons produced to the device’s volume.
\begin{figure*}[ht!]
    \centering
    \begin{subfigure}[b]{0.33\textwidth}
        \centering
        \includegraphics[width=\textwidth]{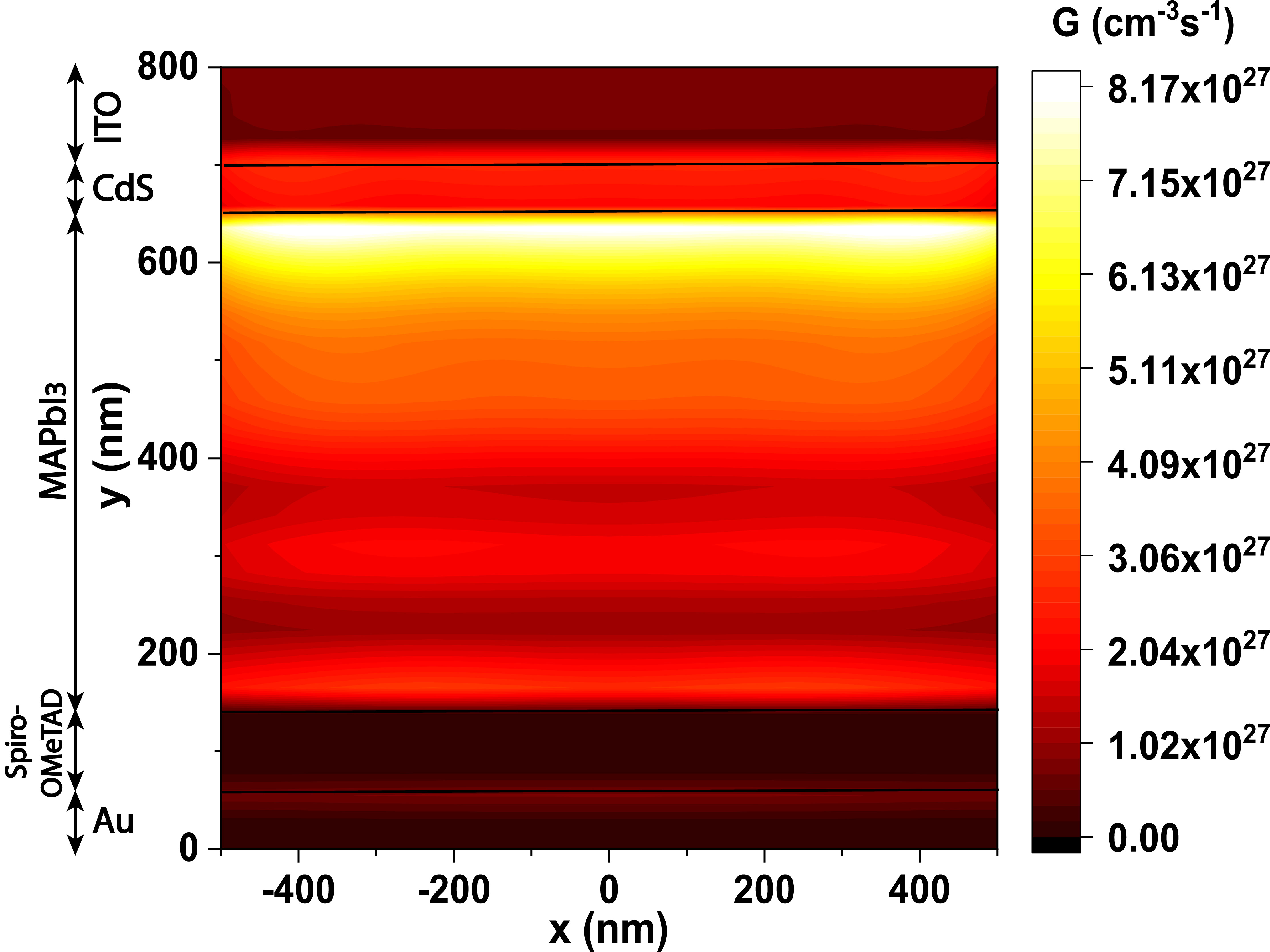}
        \caption{}
       \label{gen_planer}
    \end{subfigure}
    ~
    \begin{subfigure}[b]{0.31\textwidth}
        \centering
        \includegraphics[width=\textwidth]{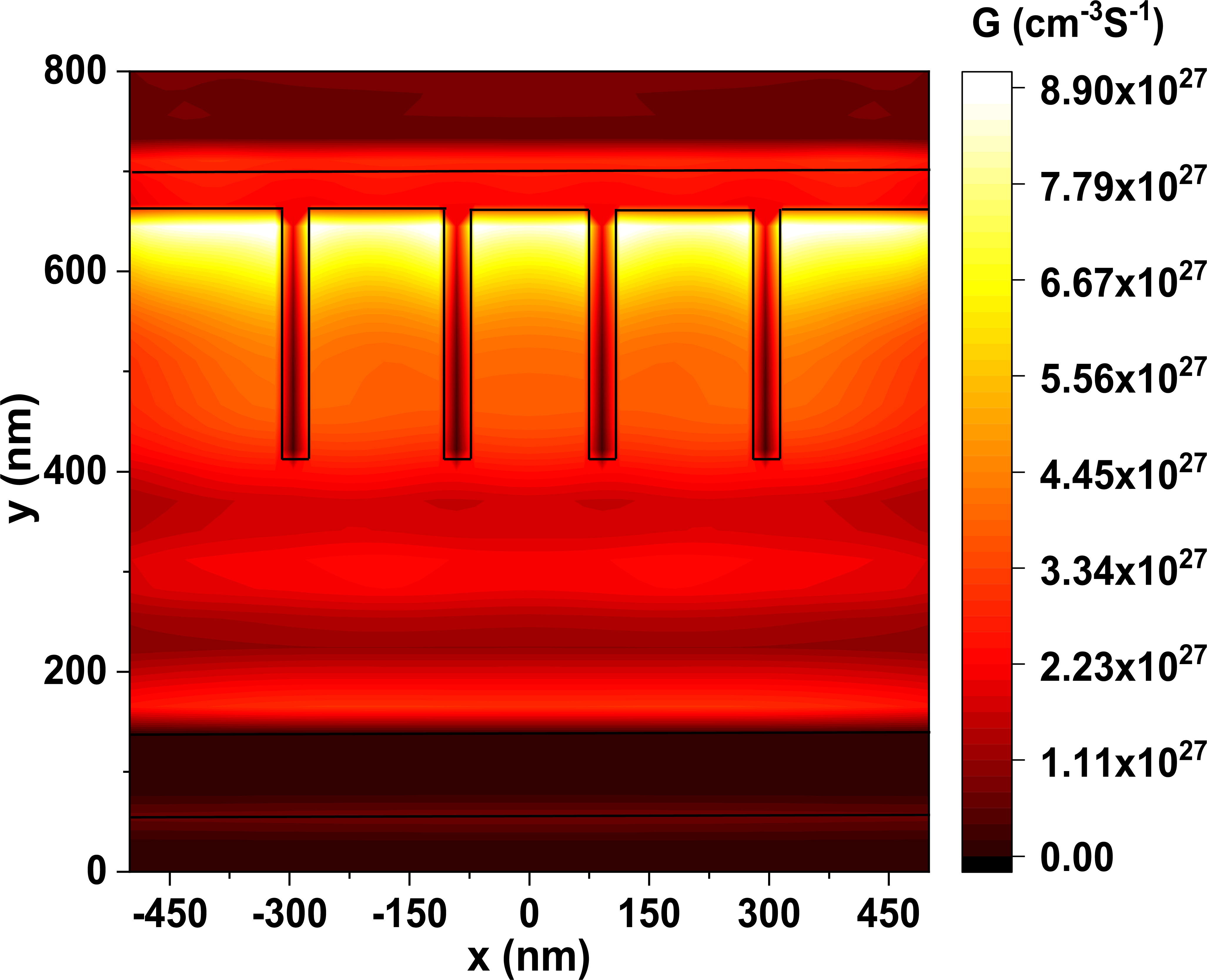}
        \caption{}
        \label{gen_nanorod}
    \end{subfigure}
   ~
    \begin{subfigure}[b]{0.31\textwidth}
        \centering
        \includegraphics[width=\textwidth]{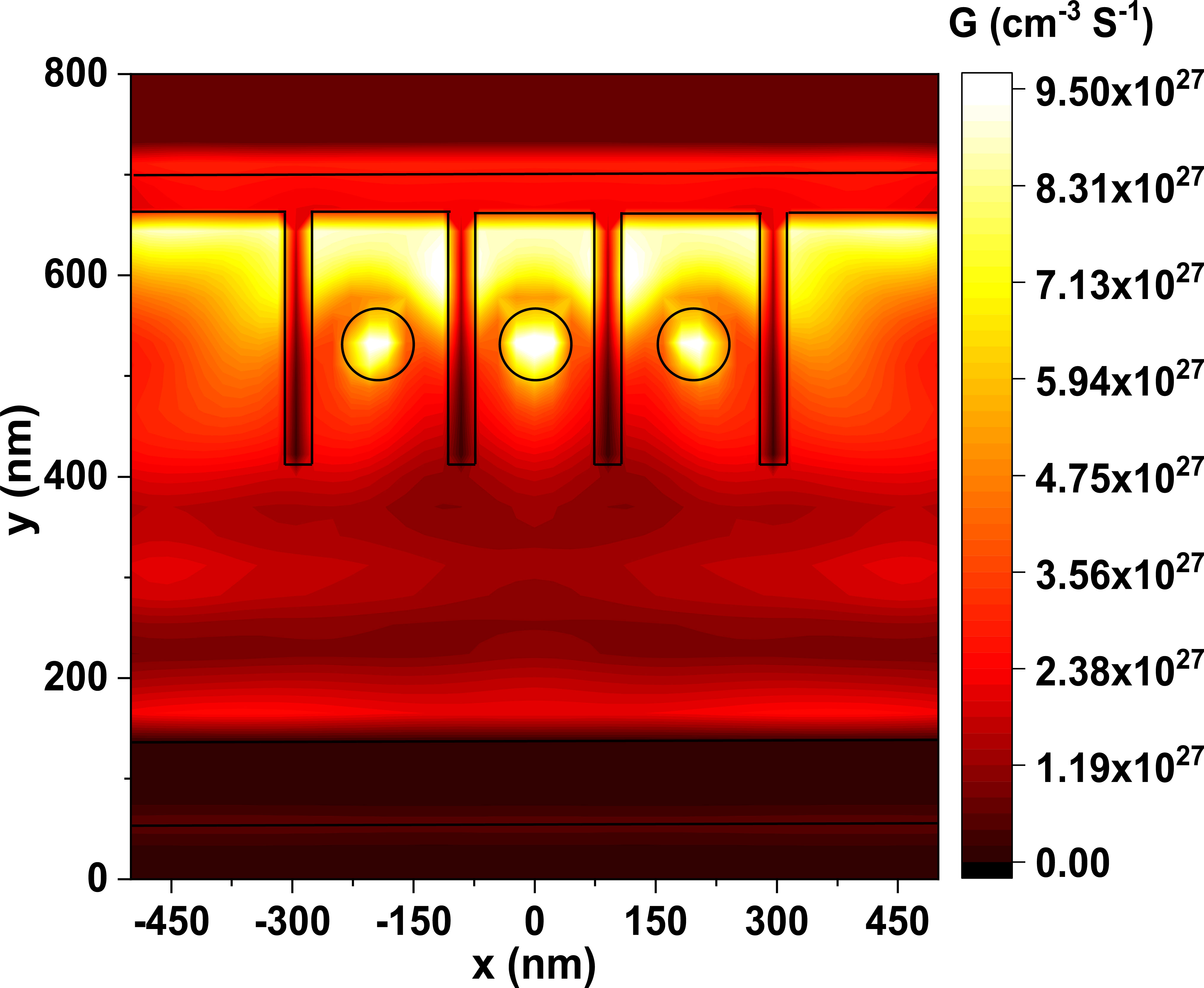}
        \caption{}
        \label{gen_gold}
    \end{subfigure}
    
    \caption{Two-dimensional generation rate profiles for three different structures of PSCs: (a) planar structure, (b) grated CdS structure, and (c) grated CdS structure with Au NPs. The color map represents the local generation rate in units of cm\textsuperscript{-3}s\textsuperscript{-1}.}
    \label{fig:gen_three_structure}
\end{figure*}

The generation rate of a solar cell can be maximized by optimizing the position of the materials in the device structure. In particular, the generation rate is highest at the interface between the absorber layer and ETL, where electron-hole pair generation is most efficient due to the substantial propensity of MAPbI\textsubscript{3} to absorb visible light. To further enhance carrier generation, plasmon stimulation of Au NPs can be employed to increase absorption in the IR region. As shown in Fig. \ref{fig:gen_main}, the addition of Au NPs results in an improvement of the generation rate curve. Figure \ref{fig:gen_three_structure} displays the 2D representation of the solar cell with the generation rate distributions, and as seen in Fig. \ref{fig:ABS_gen_main} and \ref{gen_gold}, grated CdS with Au NPs exhibits the highest generation rate of $\mathrm{1.07\times10^{28}}$ cm\textsuperscript{-3}s\textsuperscript{-1}. This is due to the increased absorption of light near the plasmon resonance of the Au NPs, resulting in a higher generation rate around the gold nanoparticles, as seen in Fig. \ref{gen_gold}.
\begin{figure}[!ht]
    \centering
    \includegraphics[width=8.5cm]{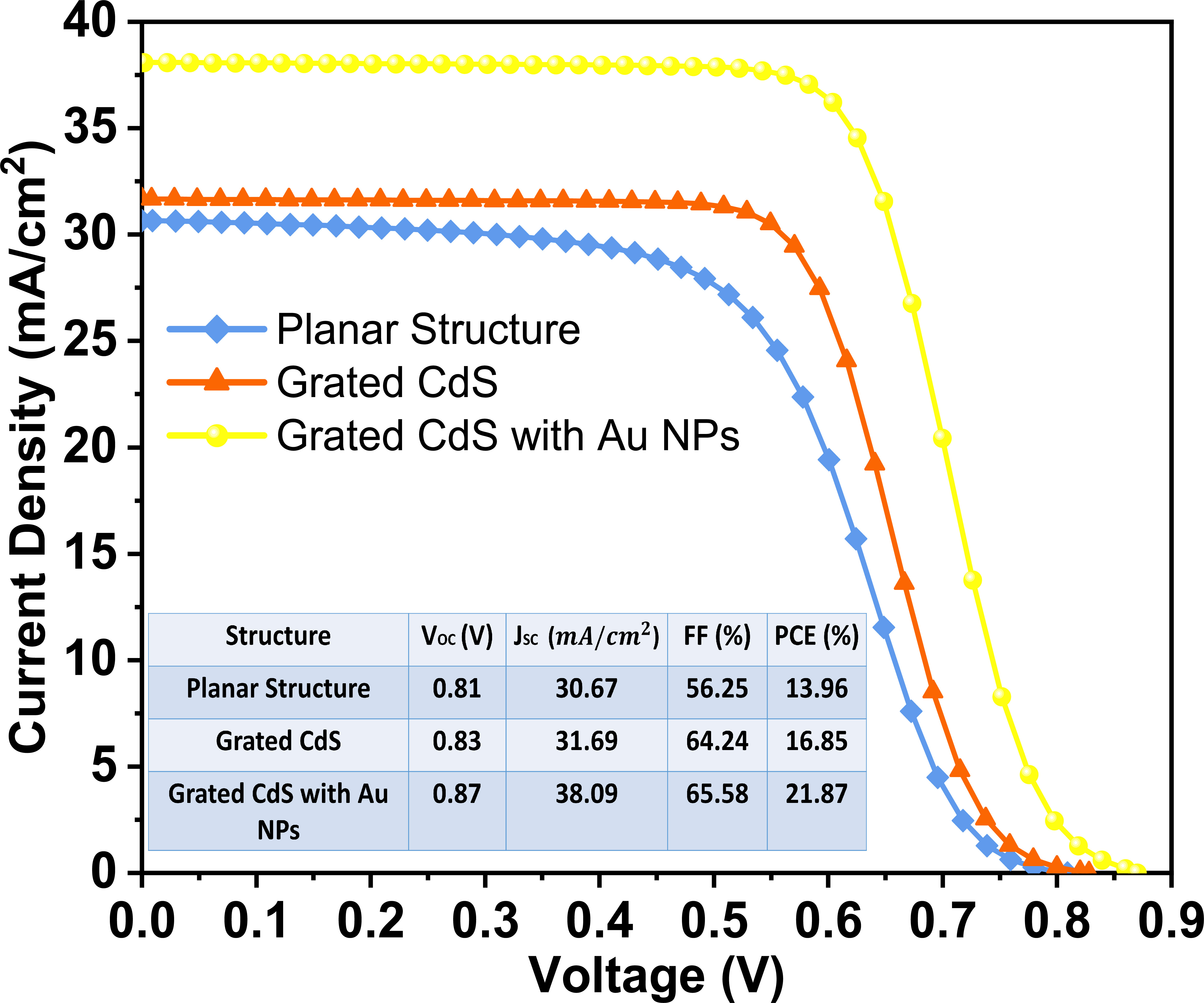}
    \caption{IV characteristics of planar, grated CdS, and grated CdS with Au NPs structures. The table inside shows the V\textsubscript{oc}, J\textsubscript{sc}, FF, and PCE for each of the structures. }
    \label{fig:IV_curve_3design}
\end{figure}

Figure \ref{fig:IV_curve_3design} shows the I-V characteristics of the three PSC structures, along with a table that provides V\textsubscript{oc}, J\textsubscript{sc}, FF, and PCE for all three structures. The grated CdS with Au NPs structure displays the highest values for all four parameters, with a substantial increase in both J\textsubscript{sc} and PCE, which climb from 30.67 mA/cm\textsuperscript{2} to 38.09 mA/cm\textsuperscript{2} and from 13.96\% to 21.87\%, respectively. Furthermore, compared to the initial planar structure, the fill factor (FF) value increases significantly from 56.25\% to 65.58\% for the final structure.

\subsection{Electric Field Profile}
\begin{figure}[!ht]
    \centering
    \includegraphics[width=14cm]{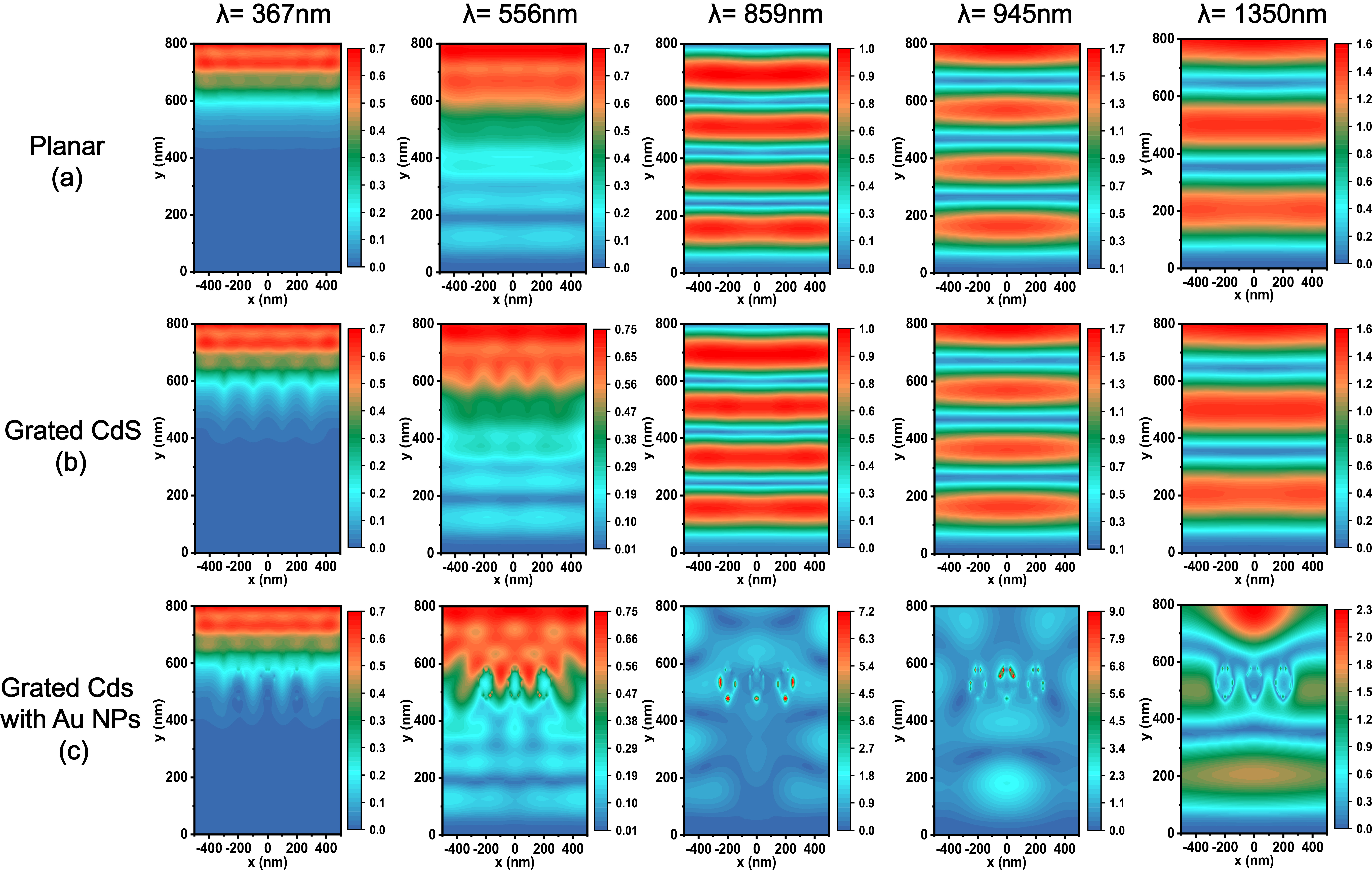}
    \caption{Electric field (E) intensity profiles for three different PSC structures at five different wavelengths: 357 nm, 556 nm, 859 nm, 945 nm, and 1350 nm. (a) For the planar structure, the maximum E field intensity is seen near the top layer of PSC when the wavelength of the light is minimal due to the low penetration depth for a short wavelength of light and the high absorption coefficient of the ETL layer. A standing wave is created within the perovskite layer due to the Fabry-Perrot resonance caused by the gold back reflector in the absence of a CdS grating and plasmonic NPs. (b) For the grated CdS structure, the absorption of PSC is identical to that of a planar structure for incoming light wavelengths greater than 750 nm. Therefore, the electric field pattern is identical to the E-field pattern of a planar structure for wavelengths of 859 nm, 945 nm, and 1350 nm. (c) For the grated CdS with Au NPs structure, due to the high absorption capacity of plasmonic Au NPs, a significantly high E field is observed around the Au NPs at 859 nm and 945 nm, while the electric field distribution is unchanged compared to the previous grated CdS structure at 357 nm wavelength due to the low penetration depth of shorter wavelengths of light.}
    \label{fig:E_profile_3design}
\end{figure}
To justify the absorption and generation spectra of these three structures, we simulate electric field intensity profiles at five different wavelengths: 357 nm, 556 nm, 859 nm, 945 nm, and 1350 nm, presented in Fig. \ref{fig:E_profile_3design}. As shown in Fig. \ref{fig:E_profile_3design}(a)(I-II), the maximum E field intensity is observed near the top layer of PSC when the wavelength of the light is minimal. This is because the penetration depth for a short wavelength of light is low and the ETL layer has a high absorption coefficient. In the absence of a CdS grating and plasmonic NPs, the electric field creates a standing wave within the perovskite layer due to the Fabry-Perrot resonance caused by the gold back reflector, as shown in Fig. \ref{fig:E_profile_3design}(a)(III-V).

 For wavelengths of 357 nm and 556 nm, most of the light is absorbed in the upper part of PSC, as previously mentioned. Due to this and the short wavelength of the light, the top portion of the solar cell exhibits the highest electric field intensity, as seen in Fig. \ref{fig:E_profile_3design}(b)(I-II). Upon implementation of CdS grating, the absorption of PSC is identical to that of a planar structure for incoming light wavelengths greater than 750 nm. Due to this, the electric field pattern is identical to the E-field pattern of a planar structure for wavelengths of 859 nm, 945 nm, and 1350 nm, as illustrated in Fig. \ref{fig:E_profile_3design}(b)(III-V). After introducing the Au NPs between the CdS grating, the electric field distribution remains unchanged compared to the previous grated CdS structure at 357 nm wavelength, as shown in Fig. \ref{fig:E_profile_3design}(c)(I) due to the low penetration depth of shorter wavelength of light. Other materials utilized in PSC, such as CdS, Spiro-OMeTAD, and MAPbI3, have significantly less capacity to absorb light beyond 750 nm compared to plasmonic Au NPs. Due to this, a significantly high E field is observed around the Au NPs at 859 nm and 945 nm, as illustrated in Fig. \ref{fig:E_profile_3design}(c)(III, IV).

\subsection{Effect of CdS Width on Generation Rate}

\begin{figure}[!ht]
    \centering
    \includegraphics[width=7cm]{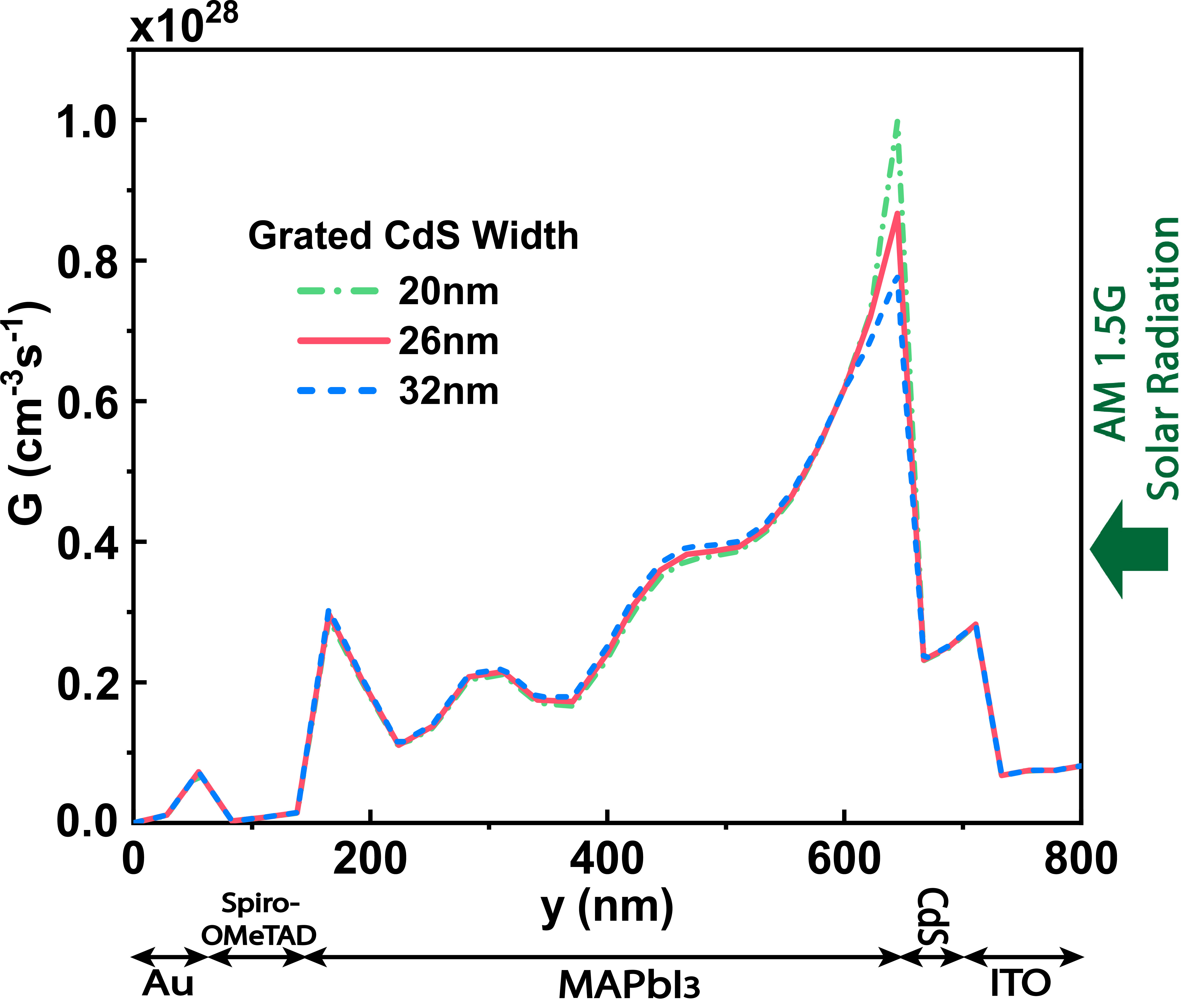}
    \caption{The generation rate as a function of depth in the PSC for CdS gratings with widths of 20 nm, 26 nm, and 32 nm. The generation rate is highest at the ETL/absorber interface due to repeated reflection of light and electron-hole pair formation, and it decreases as we move further into the active region. The interfacial area where most of the solar radiation is absorbed creating electron-hole pairs changes significantly when the grating width is altered, resulting in a very sensitive generation rate in that region. The highest generation rate of $9.98\times10^{27}$ cm\textsuperscript{-3}s\textsuperscript{-1} is observed for a grating width of 20 nm, while the lowest generation rate of $7.79\times10^{27}$ cm\textsuperscript{-3}s\textsuperscript{-1} is observed for a grating width of 32 nm.}
    \label{fig:gen_CdS_width}
\end{figure}
To maximize the generation rate, it is important to have a large interfacial area between the electron transport layer (ETL) and the absorber material, where most of the electron-hole pairs are created due to repeated reflection of light and charge separation. As shown in Fig. \ref{fig:gen_CdS_width}, the generation rate decreases as we move further into the active region because most of the incident light is absorbed in the ETL/absorber contact. To investigate the impact of grating width on the generation rate, CdS gratings with widths of 20 nm, 26 nm, and 32 nm are simulated.

As the grating width decreases, the generation rate at the interface increases due to the larger interfacial area between the ETL and absorber, where most of the solar radiation is absorbed, leading to the creation of more electron-hole pairs. The generation rate is relatively uniform for all grating widths along the Y-axis of the solar cell except for the ETL/absorber interface, which is very sensitive to changes in the interfacial area. Near the ETL/absorber junction, the highest generation rate of $9.98\times10^{27}$ cm\textsuperscript{-3}s\textsuperscript{-1} is observed for a grating width of 20 nm, while the generation rate is $7.79\times10^{27}$ cm\textsuperscript{-3}s\textsuperscript{-1} for a grating width of 32 nm.

\subsection{Gold Nanoparticle Radius optimization}
\begin{figure*}[ht!]
    \centering
    \begin{subfigure}[b]{0.32\textwidth}
        \centering
        \includegraphics[width=\textwidth]{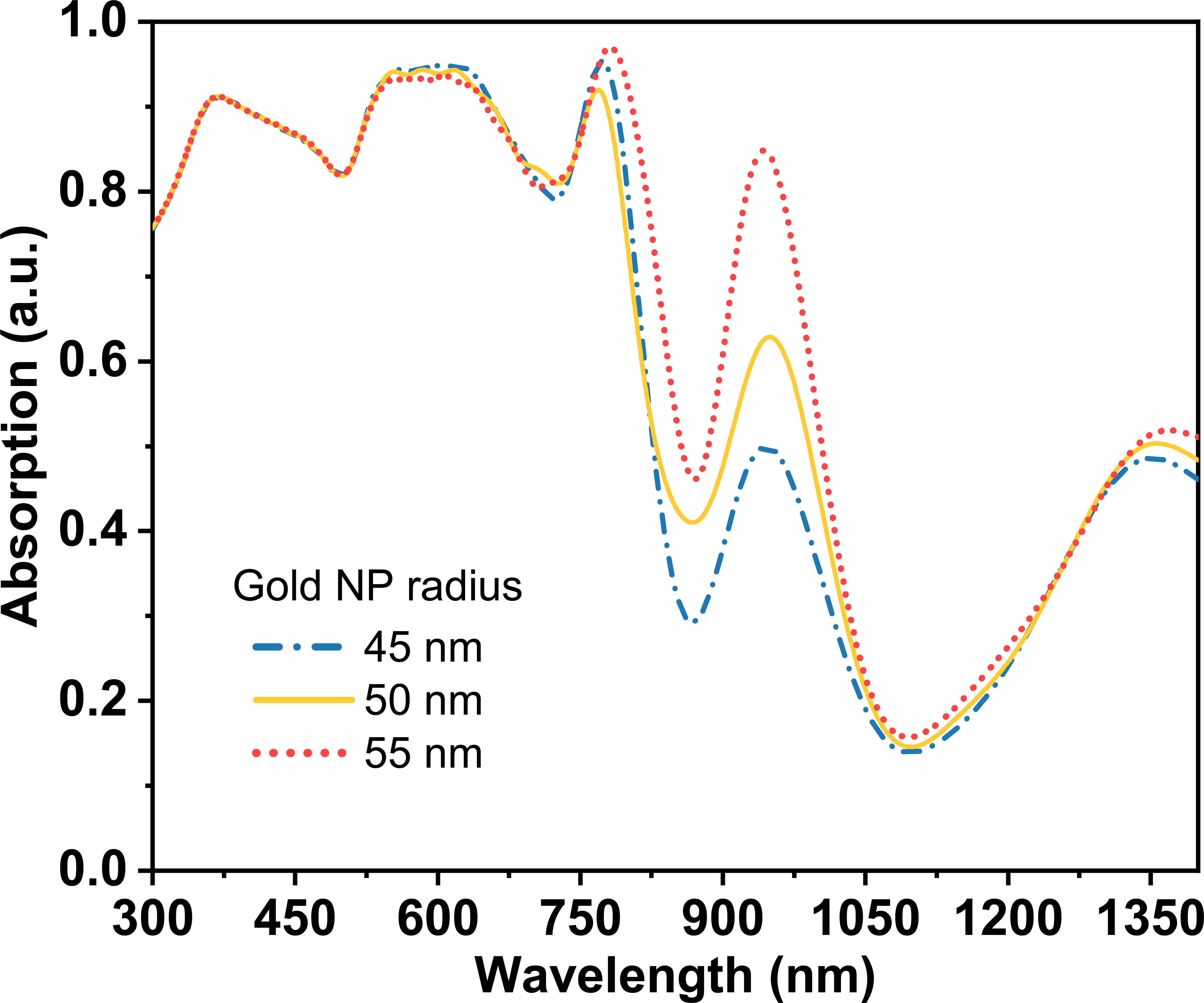}
        \caption{}
       \label{fig:abs_gold_radius}
    \end{subfigure}
    ~
    \begin{subfigure}[b]{0.325\textwidth}
        \centering
        \includegraphics[width=\textwidth]{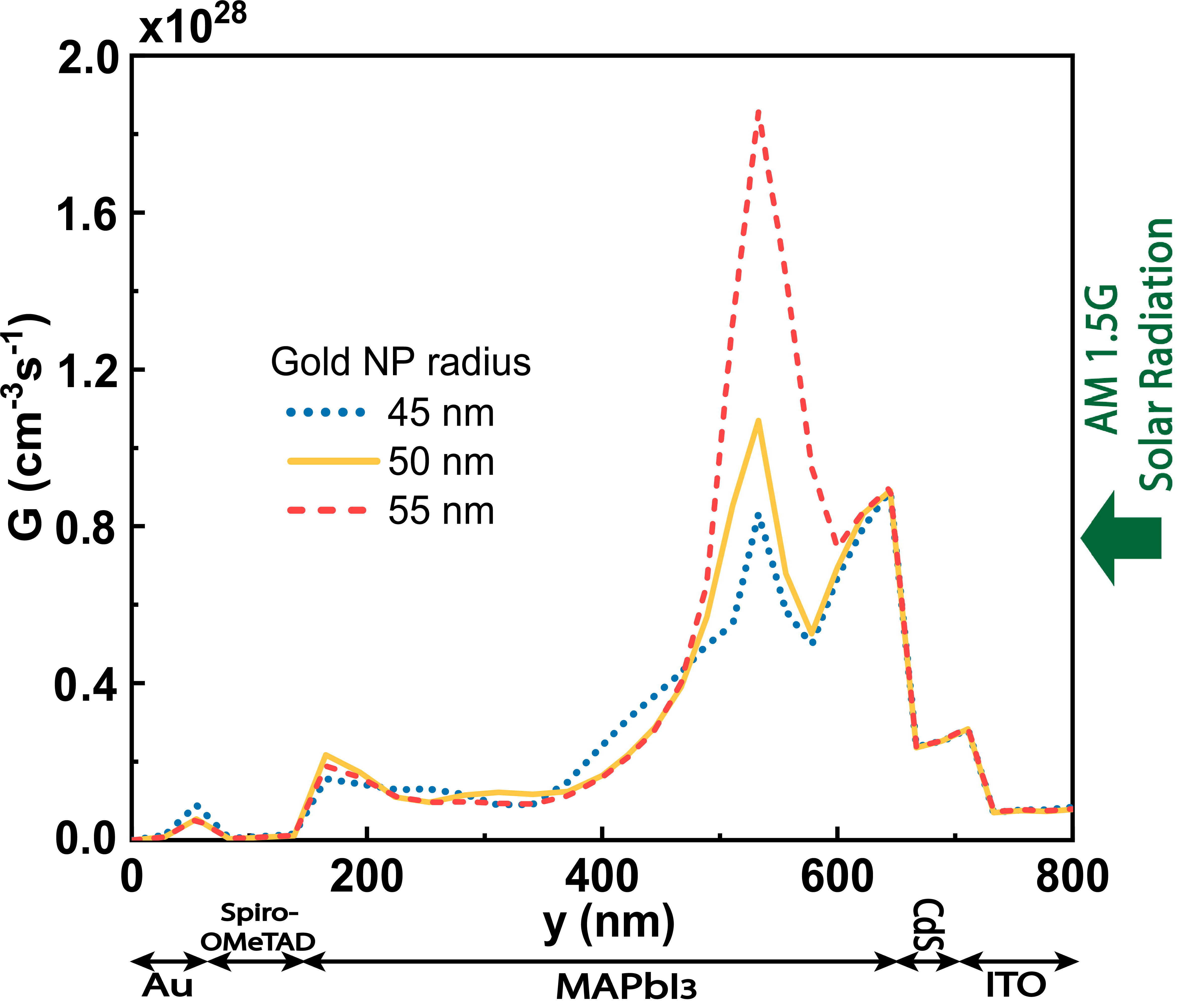}
        \caption{}
        \label{fig:gen_gold_radius}
    \end{subfigure}
   ~
    \begin{subfigure}[b]{0.32\textwidth}
        \centering
        \includegraphics[width=\textwidth]{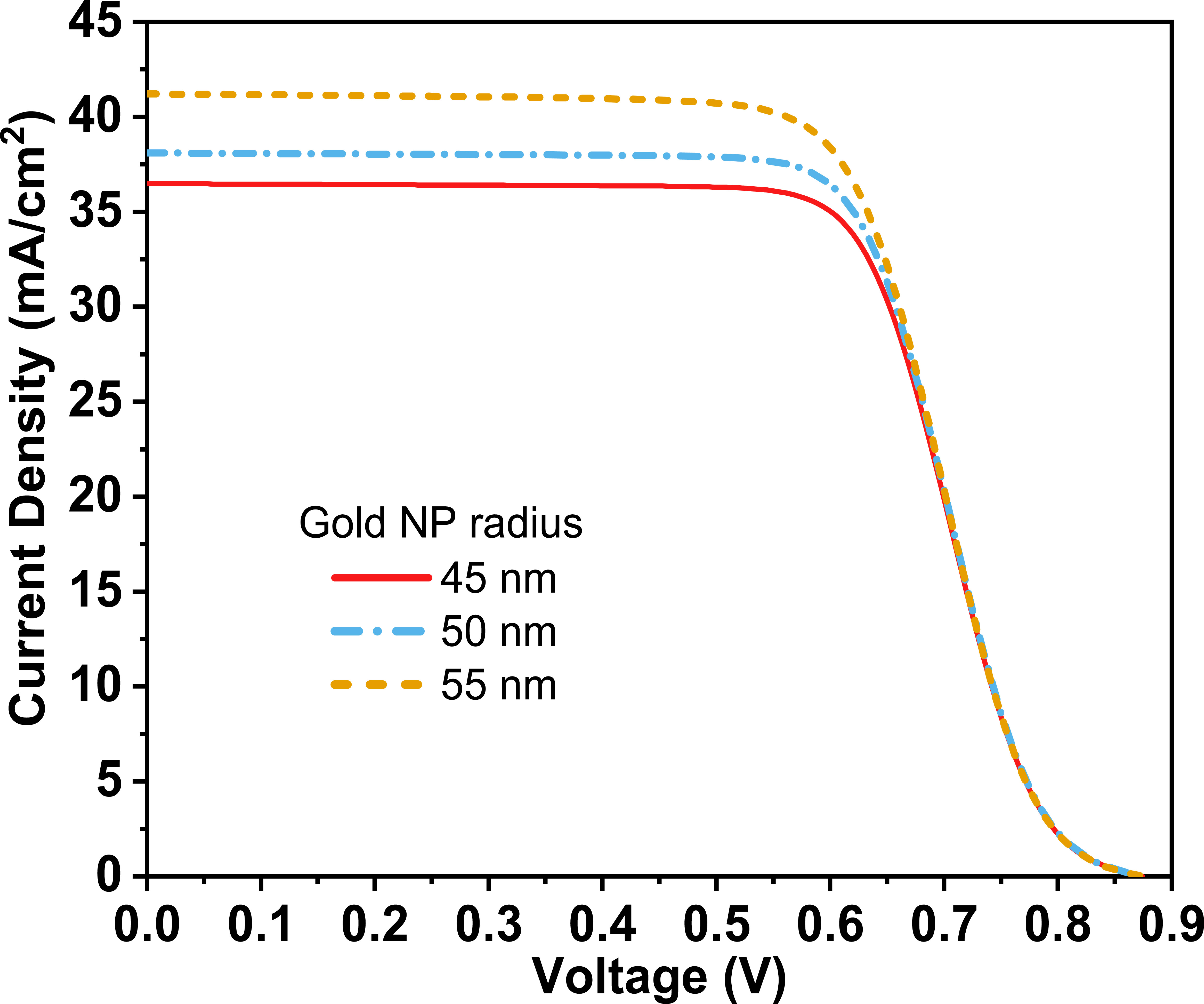}
        \caption{}
        \label{fig:IV_gold_radius}
    \end{subfigure}
    
   \caption{Effect of the gold nanoparticle (NP) radius on the (a) absorption spectra, (b) generation rate curves, and (c) current-voltage (IV) characteristics of the solar cell. The performance enhancement is observed in the near-infrared range beyond the visible range as the radius of the gold NP increases. The enhancement in the absorption and generation rate is primarily due to the localized surface plasmon resonance (LSPR) effect, which is proportional to the surface area of the NPs. In addition, the larger NPs induce stronger dipole moments at the metal/dielectric interface and scatter more light according to the Mie theory, which enhances light harvesting in the absorber layer. This leads to an increase in the short-circuit current density (J\textsubscript{sc}) in the IV characteristics curve, particularly for the 55 nm radius NPs.}
    \label{fig:ABS_gen_IV_gold_radius}
\end{figure*}

The absorption and generation rates in the visible range remain unaffected by the presence of plasmonic nanoparticles (NPs) in the solar cell structure, as shown in Fig. \ref{fig:abs_gold_radius} and Fig. \ref{fig:gen_gold_radius}. However, when gold NPs of radius 45 nm, 50 nm, and 55 nm are placed in the device, a significant performance enhancement is observed beyond the visible range, as demonstrated in Fig. \ref{fig:abs_gold_radius} through Fig. \ref{fig:IV_gold_radius}. The improvement is primarily due to the larger surface area of the NPs, which enables them to interact with more light and generate more localized surface plasmon resonances (LSPRs) than smaller NPs.

As the incident light stimulates the NPs and plasmon resonance is achieved, dipole moments arise at the metal/dielectric interface, leading to the linking and scattering of photons. According to Mie's theory, the amount of light scattering is directly proportional to the size of the NPs, which is directly related to the radius of the plasmonic NPs \cite{perrakis2019efficient}. To enhance the light-harvesting in the devices, larger NPs are preferable, as they improve the light-scattering rate and lengthen the optical path of incoming light, especially sub-wavelength light, in the absorber layer. The increased absorption and generation rates due to the larger NPs result in a higher J\textsubscript{sc} value for a radius of 55 nm, as evident from the I-V characteristics curve shown in Fig. \ref{fig:IV_gold_radius}.

\subsection{Effect of plasmonic NPs position}
In this section, we investigate the impact of the vertical location of gold nanoparticles on the optical and electrical performance of perovskite solar cells (PSCs). The depth at which nanoparticles are deposited inside the perovskite layer plays a crucial role in determining the absorption and generation rate of PSCs. To this end, we insert plasmonic nanoparticles at three distinct locations within the PSC: between grated CdS (530 nm in the Y axis), at the center of the absorber layer (320 nm in the Y axis), and at the bottom of the absorber layer (200 nm in the Y axis), and study their impact on the PSC's light absorption, generation rate, and I-V curves.

\begin{figure*}[ht!]
    \centering
    \begin{subfigure}[b]{0.32\textwidth}
        \centering
        \includegraphics[width=\textwidth]{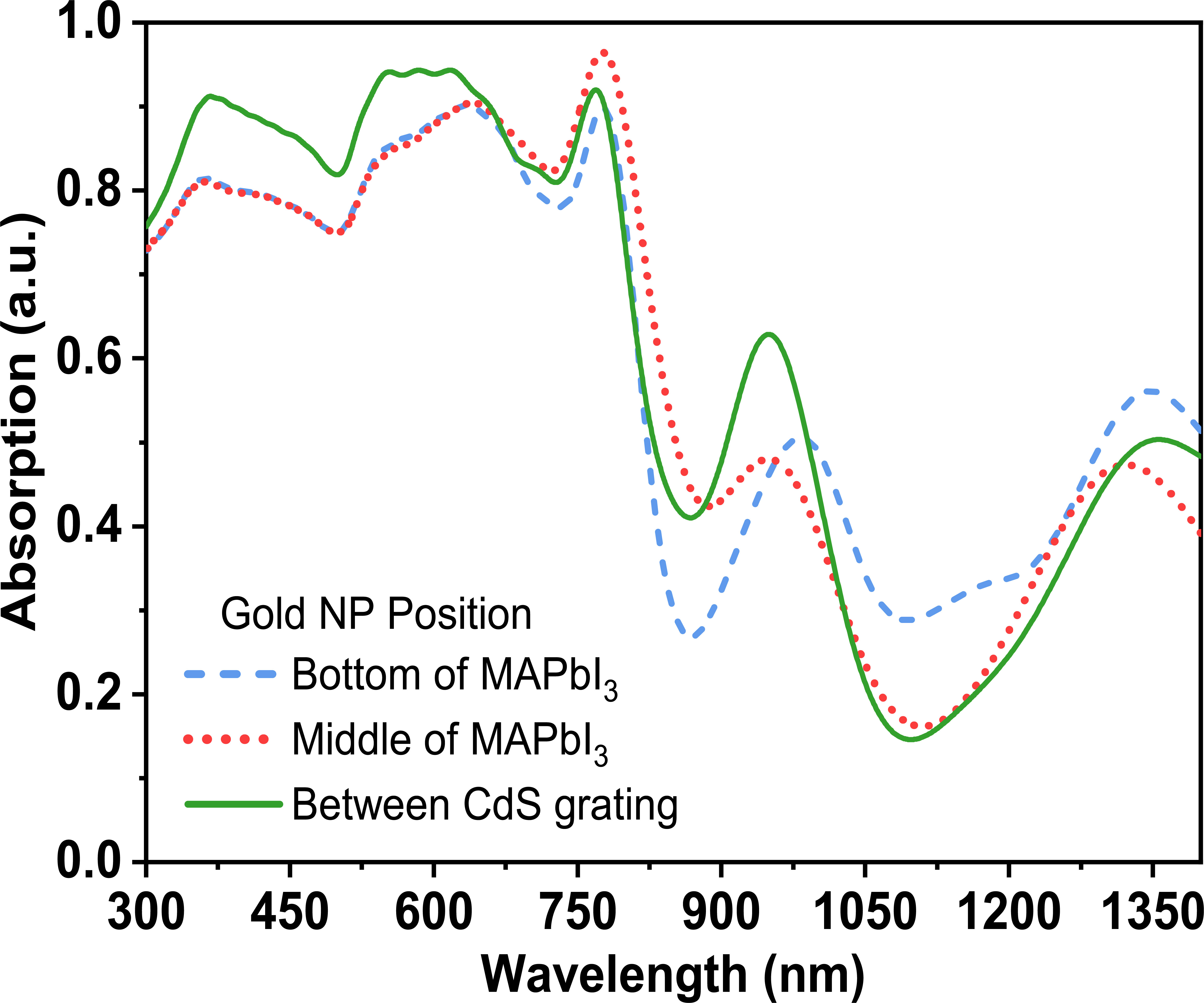}
        \caption{}
       \label{abs_gold_position}
    \end{subfigure}
    ~
    \begin{subfigure}[b]{0.33\textwidth}
        \centering
        \includegraphics[width=\textwidth]{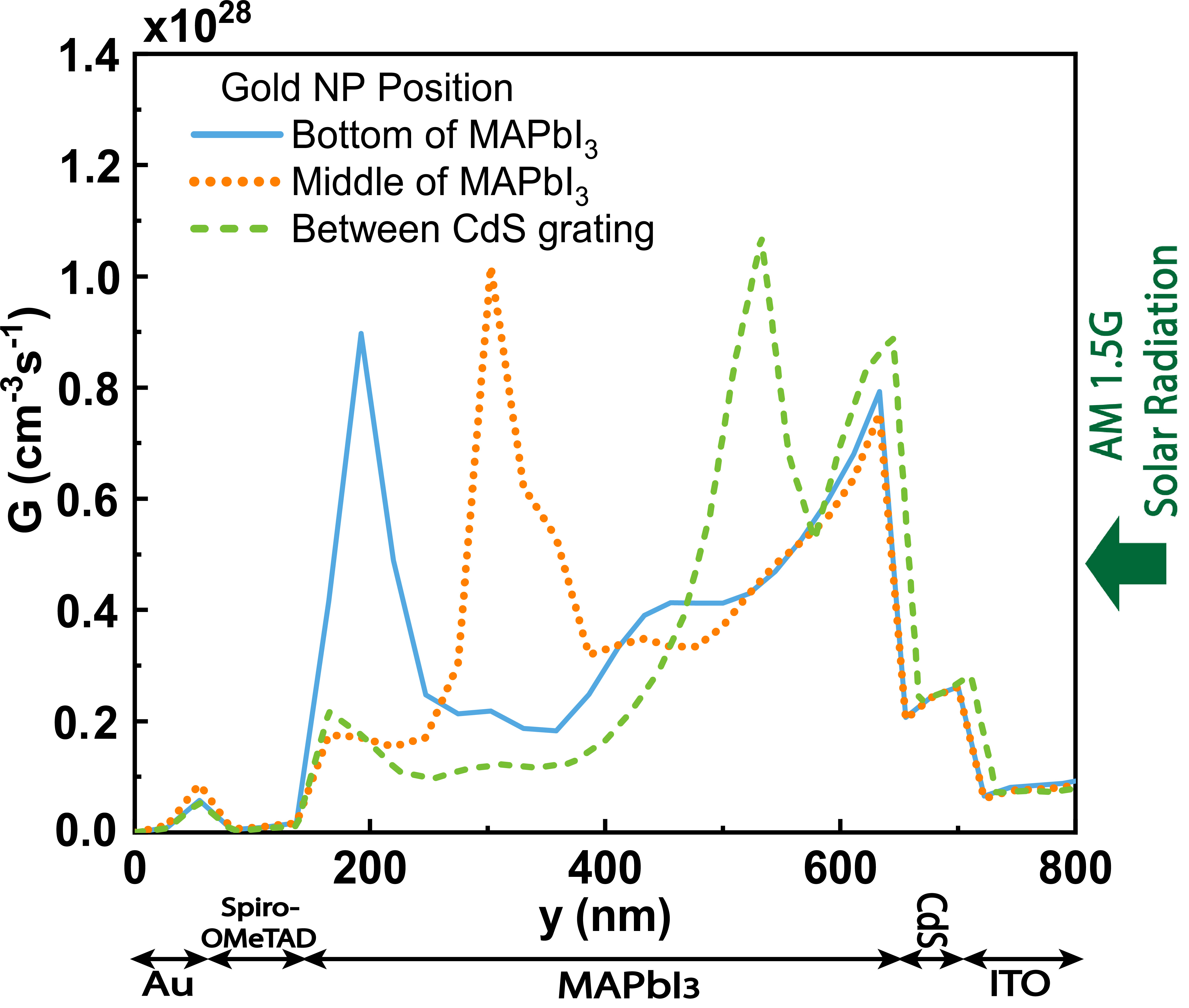}
        \caption{}
        \label{gen_gold_position}
    \end{subfigure}
   ~
    \begin{subfigure}[b]{0.32\textwidth}
        \centering
        \includegraphics[width=\textwidth]{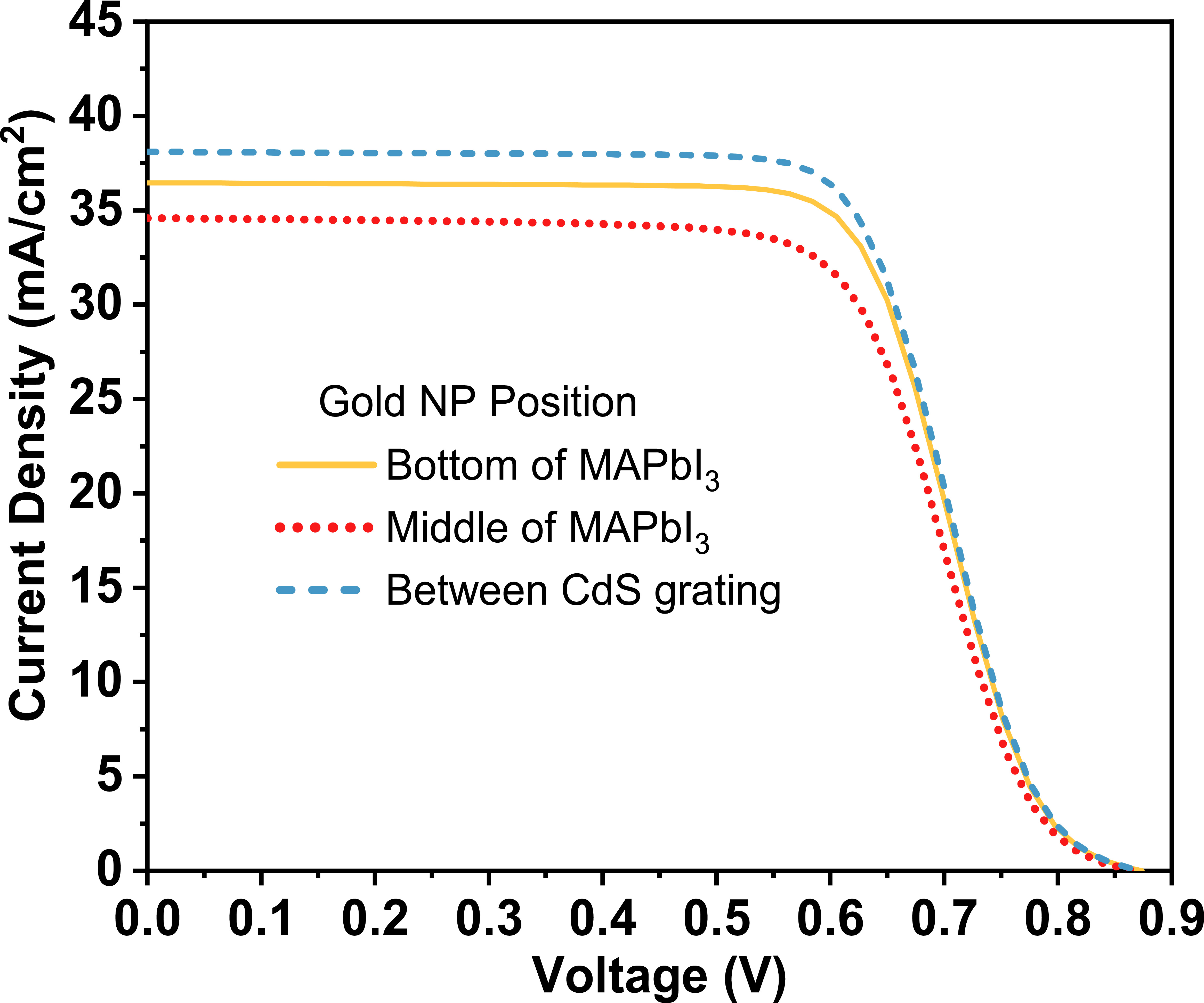}
        \caption{}
        \label{IV_gold_position}
    \end{subfigure}
    
    \caption{ The impact of the vertical location of gold nanoparticles on the performance of perovskite solar cells, as measured by (a) light absorption, (b) generation rate, and (c) I-V characteristics. Three distinct positions are investigated: between grated CdS (at 530 nm in the Y-axis), almost at the center of the absorber layer (at 320 nm in the Y-axis), and at the bottom of the absorber layer (at 200 nm in the Y-axis). The absorption spectra show that placing the gold nanoparticles between grated CdS leads to the highest overall absorption and generation rate, with a peak IR absorption at 945 nm and 62\% absorption. This position also minimizes electron-hole recombination and decreases recombination traps, leading to the highest photocurrent (38.09 mA/cm\textsuperscript{2}) and power conversion efficiency (21.87\%). }
    \label{ABS_gen_IV_gold_position}
\end{figure*}

Figure \ref{abs_gold_position} shows the absorption spectra of the PSCs as a function of the wavelength of the incident light for different vertical locations of gold NPs. When NPs are inserted between grated CdS, the PSC absorbs light most effectively in the 300-800 nm range. This location of Au NPs also creates the largest IR absorption peak at 945 nm, with 62\% absorption. Placing NPs between grated CdS ensures that most of the solar radiation is absorbed by the gold NPs rather than the MAPbI\textsubscript{3} absorber layer, which minimizes the possibility of electron-hole recombination and decreases recombination traps. The position of NPs between grated CdS also ensures that electrons can readily find a path to the cathode side through the CdS grating prior to recombination. The solar generation rate is highest at the top of the absorber layer, which helps the gold NPs to absorb more light when they are placed between CdS grating.

Figure \ref{gen_gold_position} shows the generated rate curves for the three distinct Au NP positions. The generation rate decreases monotonically from the top to the bottom of the absorber layer. The highest photocurrent and PCE values are determined to be 38.09 mA/cm\textsuperscript{2} and 21.87\%, respectively, for the gold NPs located between CdS gratings, as this position had the highest generation rate and overall absorption, which is confirmed by the I-V curves in Fig. \ref{IV_gold_position}. In contrast, the current density and PCE values for NPs in the middle of the absorber layer are 34.76 mA/cm\textsuperscript{2} and 20.06\%, respectively. For NPs at the bottom of the MAPbI\textsubscript{3} layer, photocurrent and PCE values are 36.46 mA/cm\textsuperscript{2} and 20.98\%, respectively. It should be noted that positioning plasmonic NPs close to the ETL might occasionally increase the possibility of interface transport resistance, necessitating the use of a coating layer to provide electrical separation. As our plasmonic NPs are much below the ETL, such a coating layer is not necessary for our situation.

\subsection{Effect of Plasmonic NP Shape}
\begin{figure*}[ht!]
    \centering
    \begin{subfigure}[b]{0.32\textwidth}
        \centering
        \includegraphics[width=\textwidth]{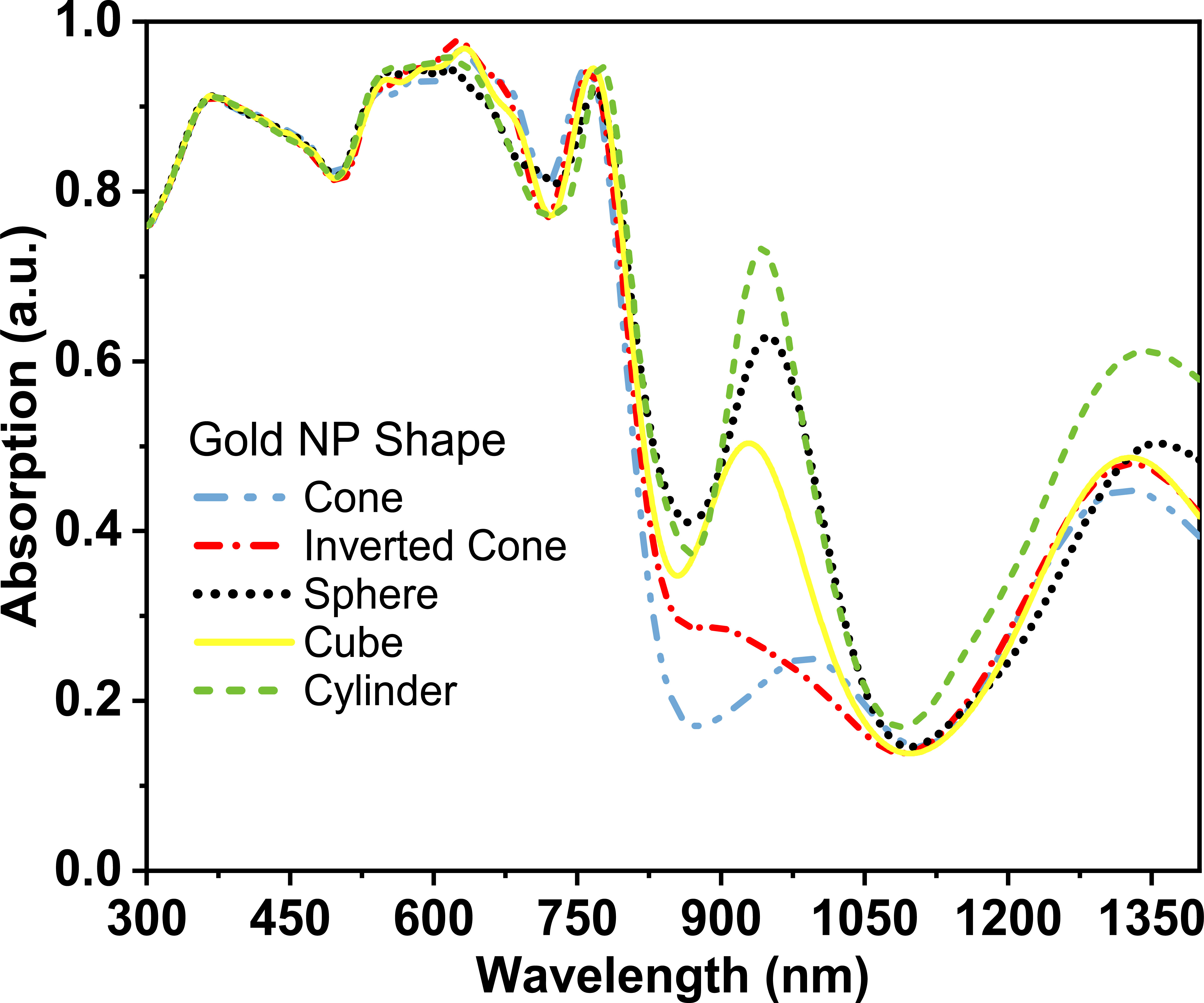}
        \caption{}
       \label{abs_gold_shape}
    \end{subfigure}
    ~
    \begin{subfigure}[b]{0.325\textwidth}
        \centering
        \includegraphics[width=\textwidth]{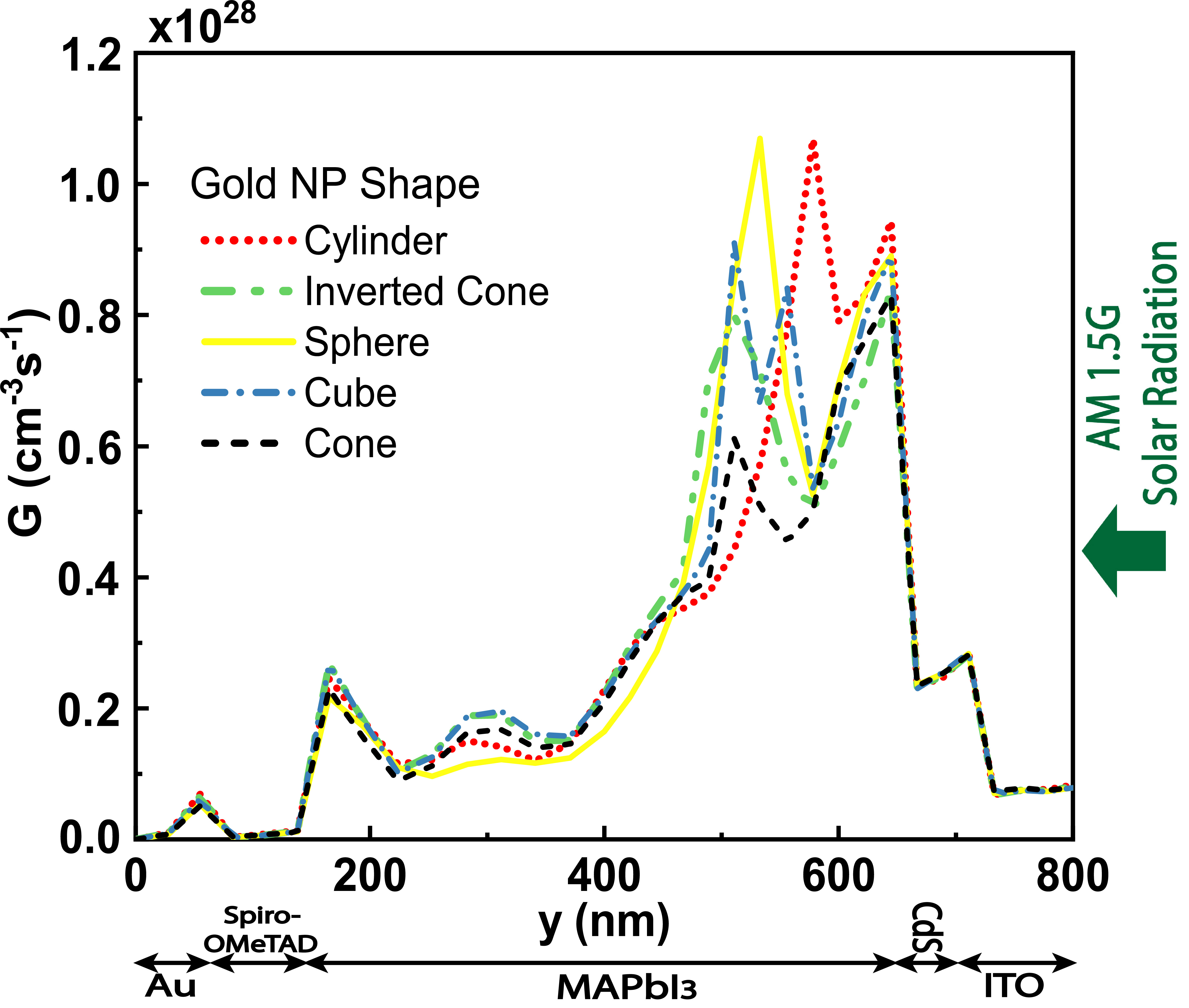}
        \caption{}
        \label{gen_gold_shape}
    \end{subfigure}
   ~
    \begin{subfigure}[b]{0.32\textwidth}
        \centering
        \includegraphics[width=\textwidth]{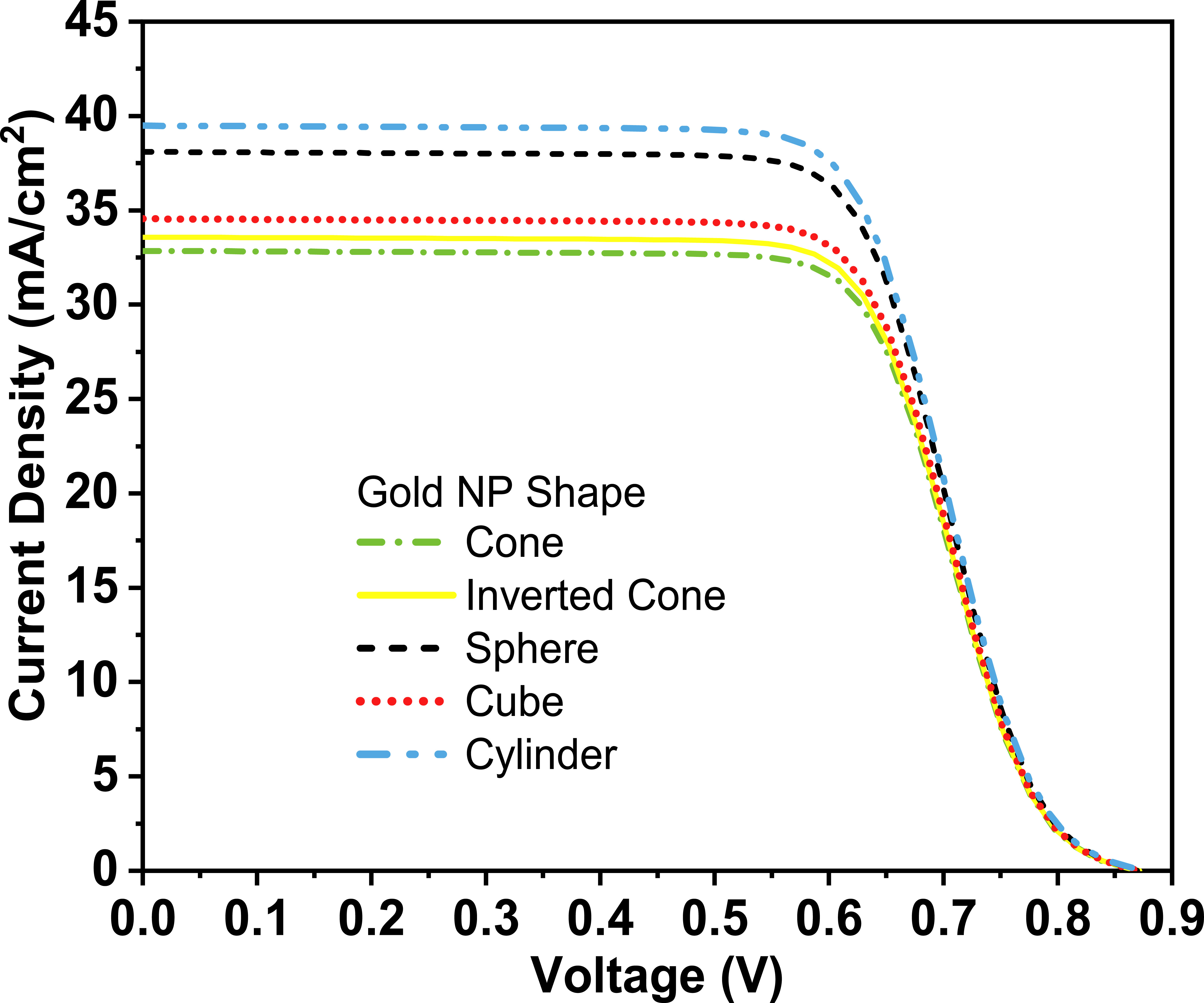}
        \caption{}
        \label{IV_gold_shape}
    \end{subfigure}
    
    \caption{Optical absorption spectra (a), generation rate curves (b), and IV characteristics (c) of a perovskite solar cell (PSC) incorporating different shapes of plasmonic nanoparticles (NPs), including cylindrical, spherical, cubic, cone, and inverted cone. The cylindrical NPs show the highest absorption enhancement and generation rate, resulting in the highest current density and power conversion efficiency (PCE) among the tested shapes. The spherical NPs exhibit the second-highest current density and a comparable PCE to cylindrical NPs, making them a more practical choice for fabrication.}
    \label{ABS_gen_IV_gold_shape}
\end{figure*}
We evaluated the optical and electrical performance of perovskite solar cells (PSCs) by implanting various metallic nanoparticles (NPs) with different shapes within the perovskite layer between CdS grating. The metallic NPs create near- and scattering effects with a plasmonic effect, enhancing PSC's absorption at longer wavelengths, and thus, its performance. To examine the augmentation of absorption in the infrared (IR) region, we simulate several nanostructures, including a cube, cone, inverted cone, and cylinder, in addition to the nanosphere. The interface area is kept nearly the same for each shape.

Figure \ref{abs_gold_shape} compares the absorption spectra of different nanostructure shapes, having the same spacing of 200 nm and volume. The cylindrical shape offers the most boost of absorption in the IR region, while the cone and inverted cone shapes show the least absorption in the IR region. The generation rate curves for various plasmonic NP shapes throughout the PSC are depicted in Fig. \ref{gen_gold_shape}. Every plasmonic NP geometry has a maximum generation rate near the top of the absorber layer, with cylinder and sphere shapes having the greatest generation rates.

Since the generation rate and absorption are both at their highest for NPs with a cylindrical form, the current density and power conversion efficiency (PCE) of the solar cell are also at their highest for this shape, with values of 39.48 mA/cm\textsuperscript{2} and 21.96\%, respectively. The second-highest current is observed at 38.09 mA/cm\textsuperscript{2} for spherical shape NPs, with a nearly identical PCE of 21.87\% to cylindrical shape NPs. Although it is easier to fabricate spherical-shaped Au NPs than cylindrical-shaped ones, the use of cylindrical-shaped NPs is recommended for PSCs. A detailed comparison of these NP shapes in terms of the highest generation rate, absorption enhancement, and electrical characteristics is provided in Fig. \ref{tab:gold_shape}.

\renewcommand{\arraystretch}{1.1}
\begin{table}[!ht]
    \small
    \centering
    \caption{Comparative analysis of inserting different gold nanoparticles shape.}
    \label{tab:gold_shape}
    \begin{tabular}{>{\centering\arraybackslash}m{4cm} >{\centering\arraybackslash}m{1.6cm} >{\centering\arraybackslash}m{1.6cm} >{\centering\arraybackslash}m{1.6cm} >{\centering\arraybackslash}m{1.6cm} >{\centering\arraybackslash}m{1.6cm}}
    \hline
        \textbf{Shape} & \includegraphics[ width=6mm]{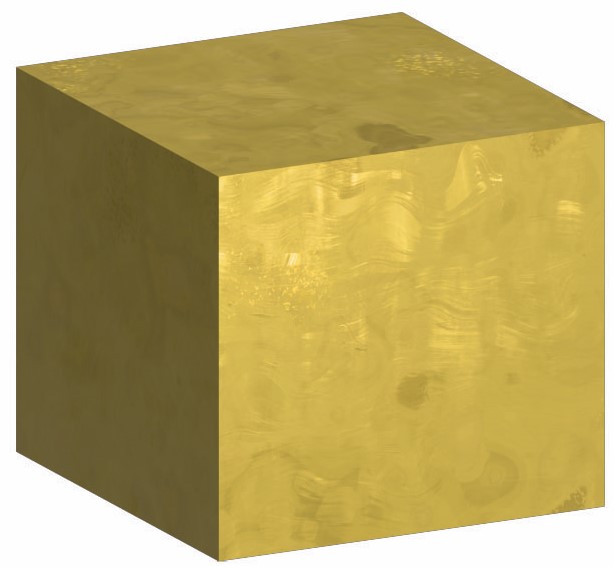}
         & \includegraphics[ width=5mm]{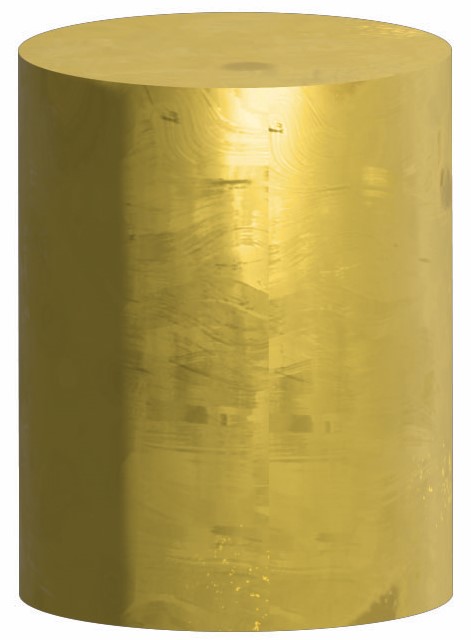}
         & \includegraphics[ width=4.8mm]{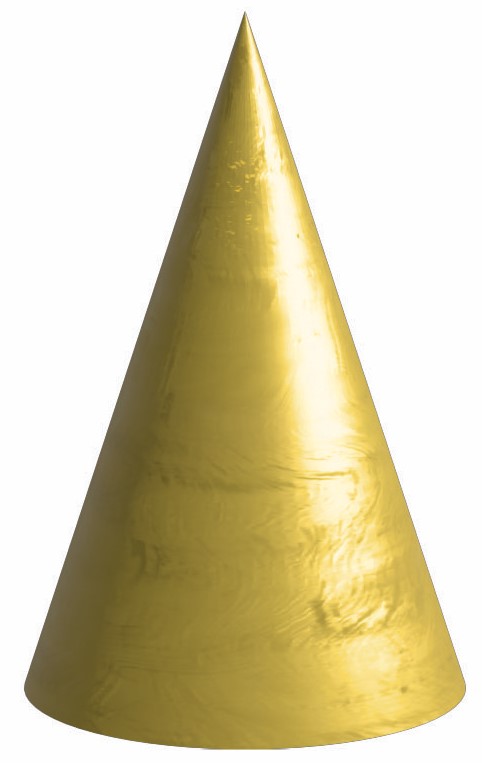}
         & \includegraphics[ height=7mm]{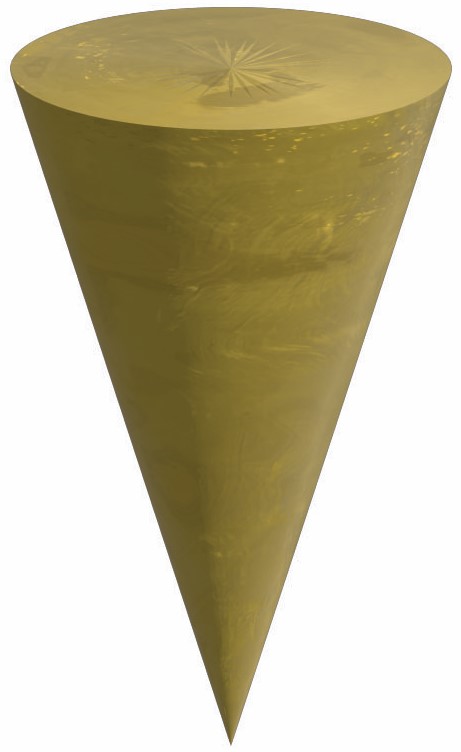}
         & \includegraphics[ width=6mm]{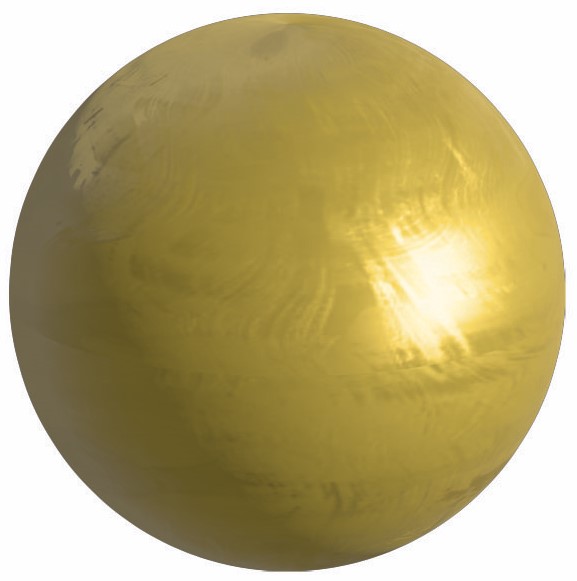}
         \\
     \hline
    Maximum Generation Rate
    (cm\textsuperscript{-3}s\textsuperscript{-1}) &	9.09 × 10\textsuperscript{27} & 1.07 × 10\textsuperscript{28} & 8.31 × 10\textsuperscript{27} & 8.33 × 10\textsuperscript{27} & 1.07 × 10\textsuperscript{28}\\
    Absorption Enhancement (\%) & 14.7 & 24.20 & 7.20 & 9.90 & 18.22 \\
    Current Density (mA/cm\textsuperscript{2}) & 34.55 & 39.48 & 32.85 & 33.58 & 38.09\\
    PCE (\%) & 18.11 & 21.96 & 17.04 & 17.15 & 21.87\\
    Fabrication Process & Moderate & Moderate & Difficult & Difficult & Easy\\
    \hline

    \end{tabular}
    
\end{table}

\subsection{Heat and thermal analysis}
\begin{figure*}[ht!]
    \centering
    \begin{subfigure}[b]{0.33\textwidth}
        \centering
        \includegraphics[width=\textwidth]{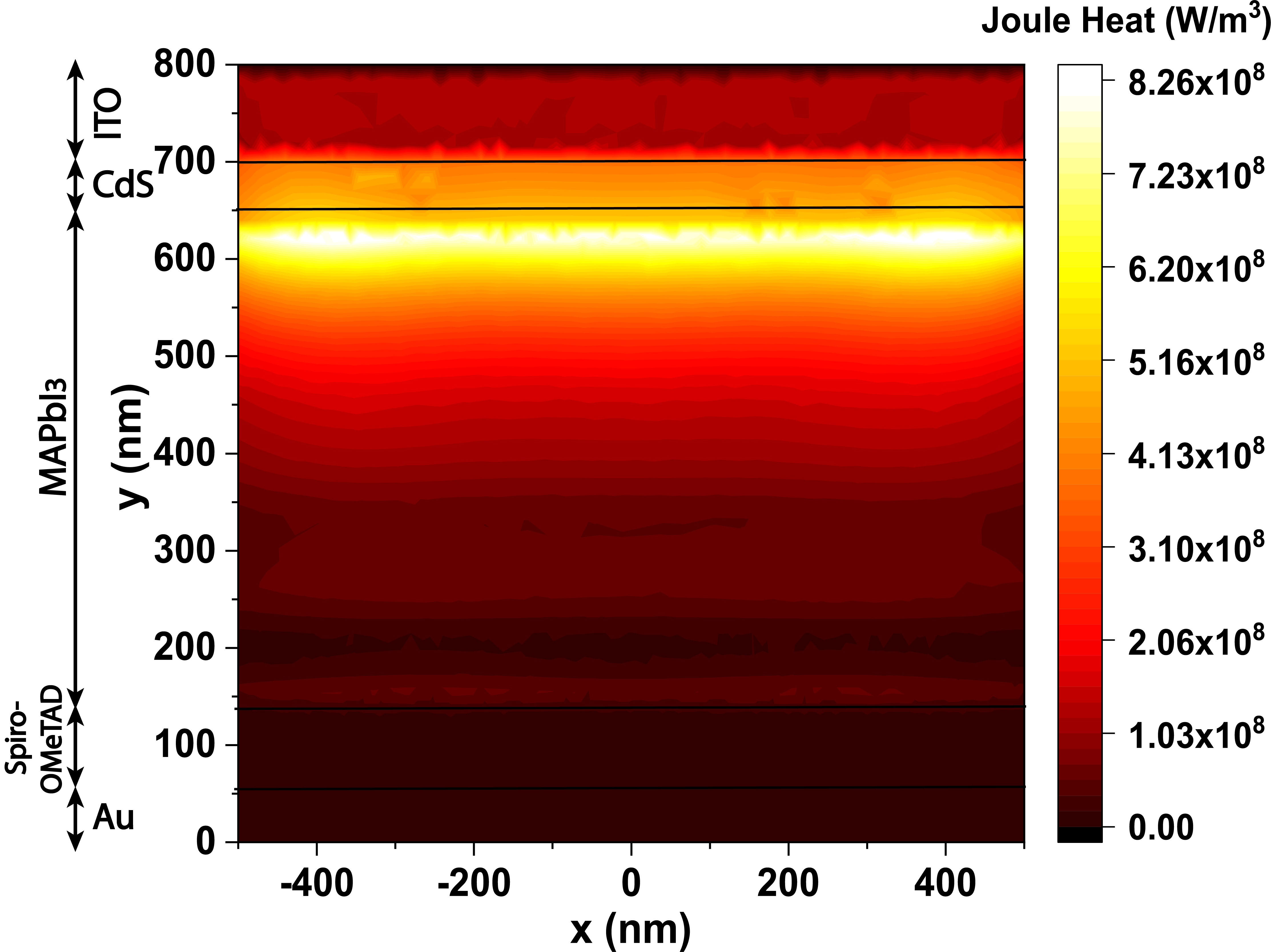}
        \caption{}
       \label{fig:Q_planer}
    \end{subfigure}
    ~
    \begin{subfigure}[b]{0.31\textwidth}
        \centering
        \includegraphics[width=\textwidth]{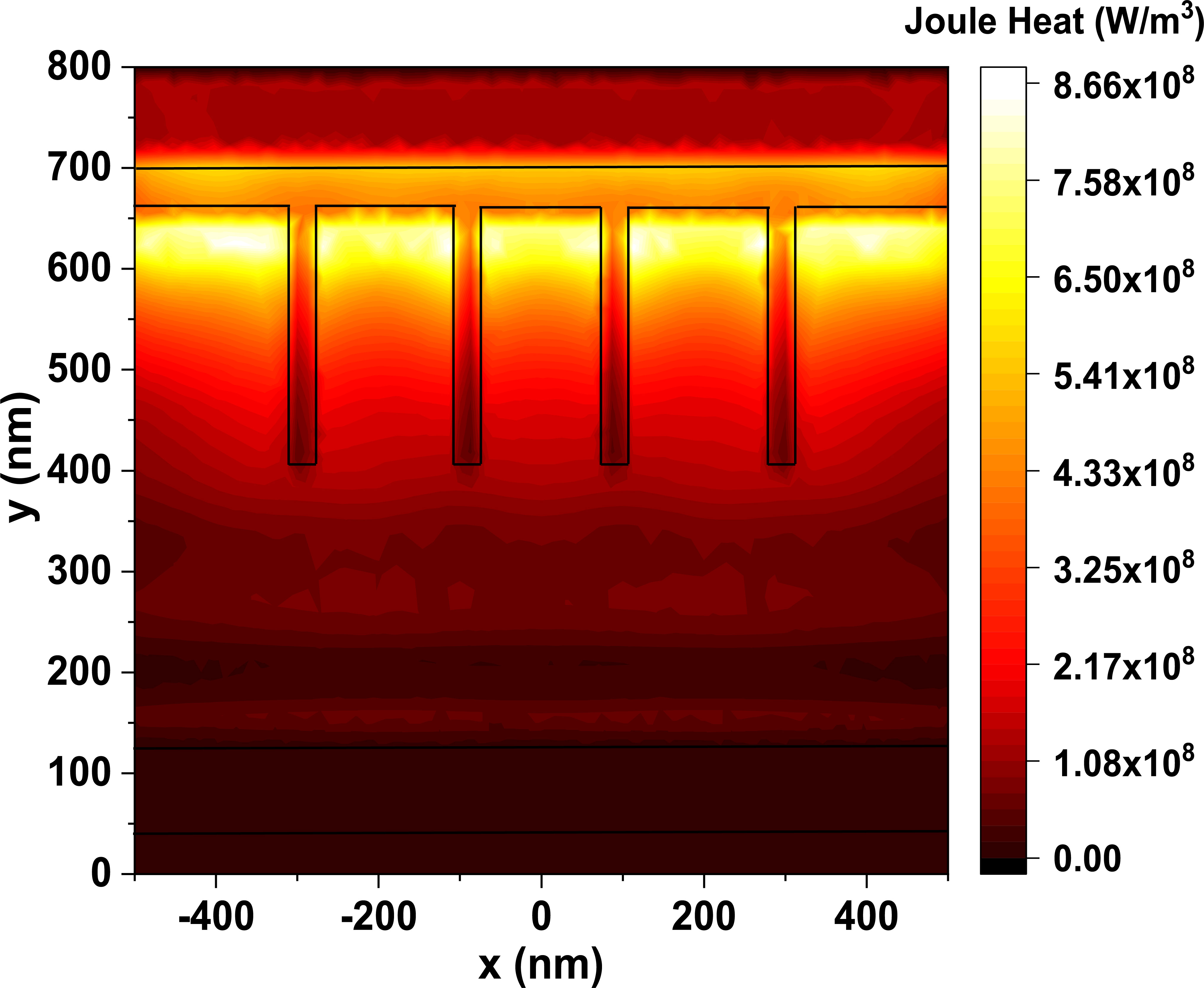}
        \caption{}
        \label{fig:Q_nanorod}
    \end{subfigure}
   ~
    \begin{subfigure}[b]{0.31\textwidth}
        \centering
        \includegraphics[width=\textwidth]{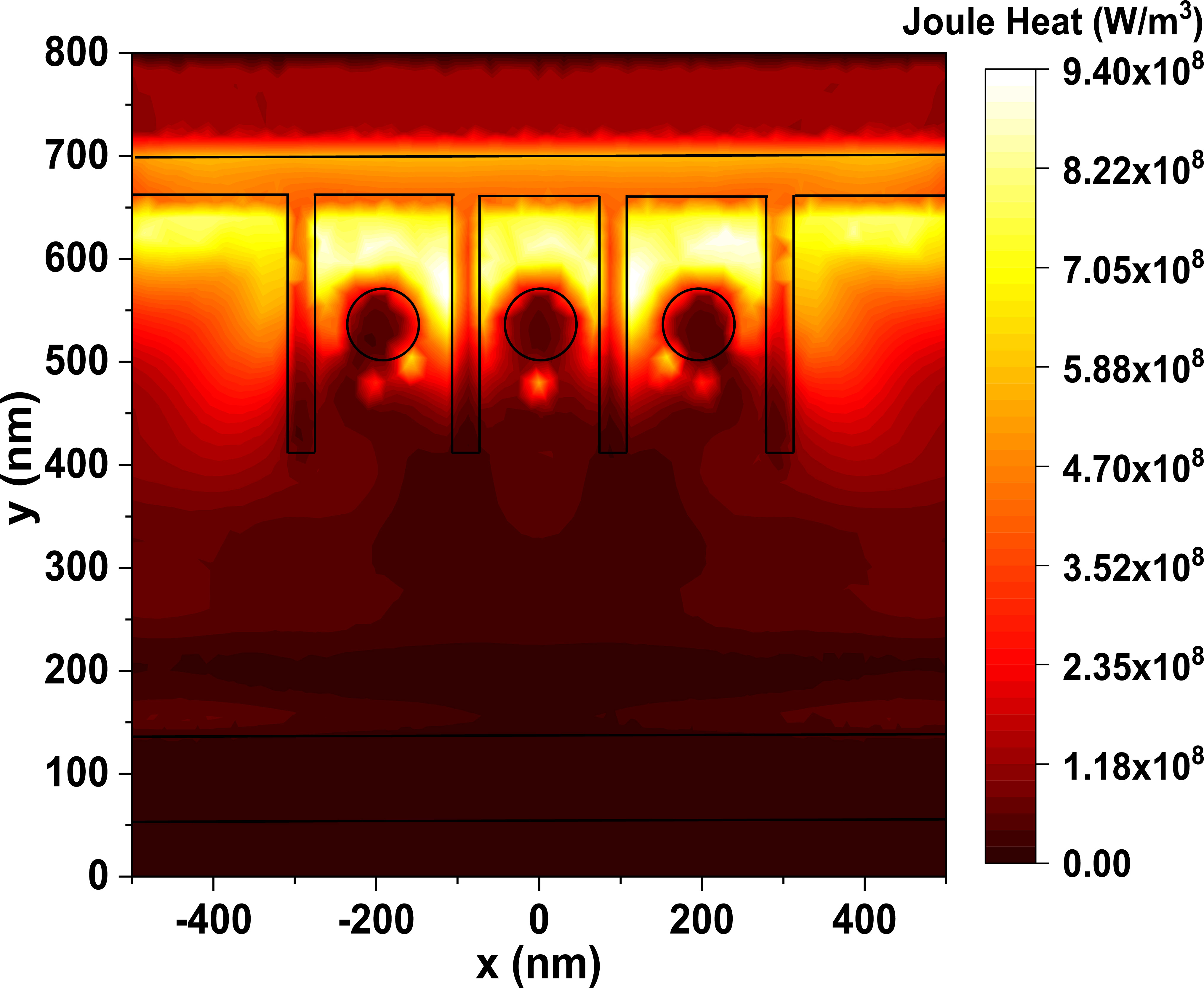}
        \caption{}
        \label{fig:Q_gold}
    \end{subfigure}
    
    \caption{The spatial distribution of non-radiative heat generated by three different solar cell structures: (a) planar solar cell, (b) grated CdS structure, and (c) grated CdS with Au plasmonic NPs. The thermal module is linked to the optic module of the simulator to determine the distribution of non-radiative heat generated in each layer of the solar cell structures and the resulting temperature gradient.}
    \label{fig:Q_3struc}
\end{figure*}
A thermal simulation is an essential tool for understanding the distribution of generated non-radiative heat and temperature gradients in different layers of a solar cell structure. The simulation results also reveal the impact of plasmonics on heat generation and temperature distribution within the solar cell structure. Linking the thermal module to the optic module of the simulator allows for thermal simulation to determine the distribution of non-radiative heat generated in the solar cell structures and the temperature gradient in each layer. Figure \ref{fig:Q_3struc} illustrates the spatial distribution of non-radiative heat generated by the three different structures. The planar solar cell, as shown in Fig. \ref{fig:Q_planer}, exhibits maximum heating at the CdS and perovskite surfaces due to the highest SRH recombination rate \cite{zandi2020numerical}. Additionally, excess energy above the bandgap $(h \nu-E_g)$ generates heat through the thermalization process, which is spatially dependent on the generation rate profile \cite{shang2017optoelectronic}. The non-radiative heat distribution for the grated CdS structure is shown in Fig. \ref{fig:Q_nanorod}, with heat generation slightly higher than in the planar structure due to the higher generation rate profile. The CdS grating area has a separate low-heated zone, acting as a heat spreader deep into the active region (perovskite). Figure \ref{fig:Q_gold} shows the thermal impact of adding the Au plasmonic NPs, resulting in a heated zone near the Au NPs.

\begin{figure*}[ht!]
    \centering
    \begin{subfigure}[b]{0.34\textwidth}
        \centering
        \includegraphics[width=\textwidth]{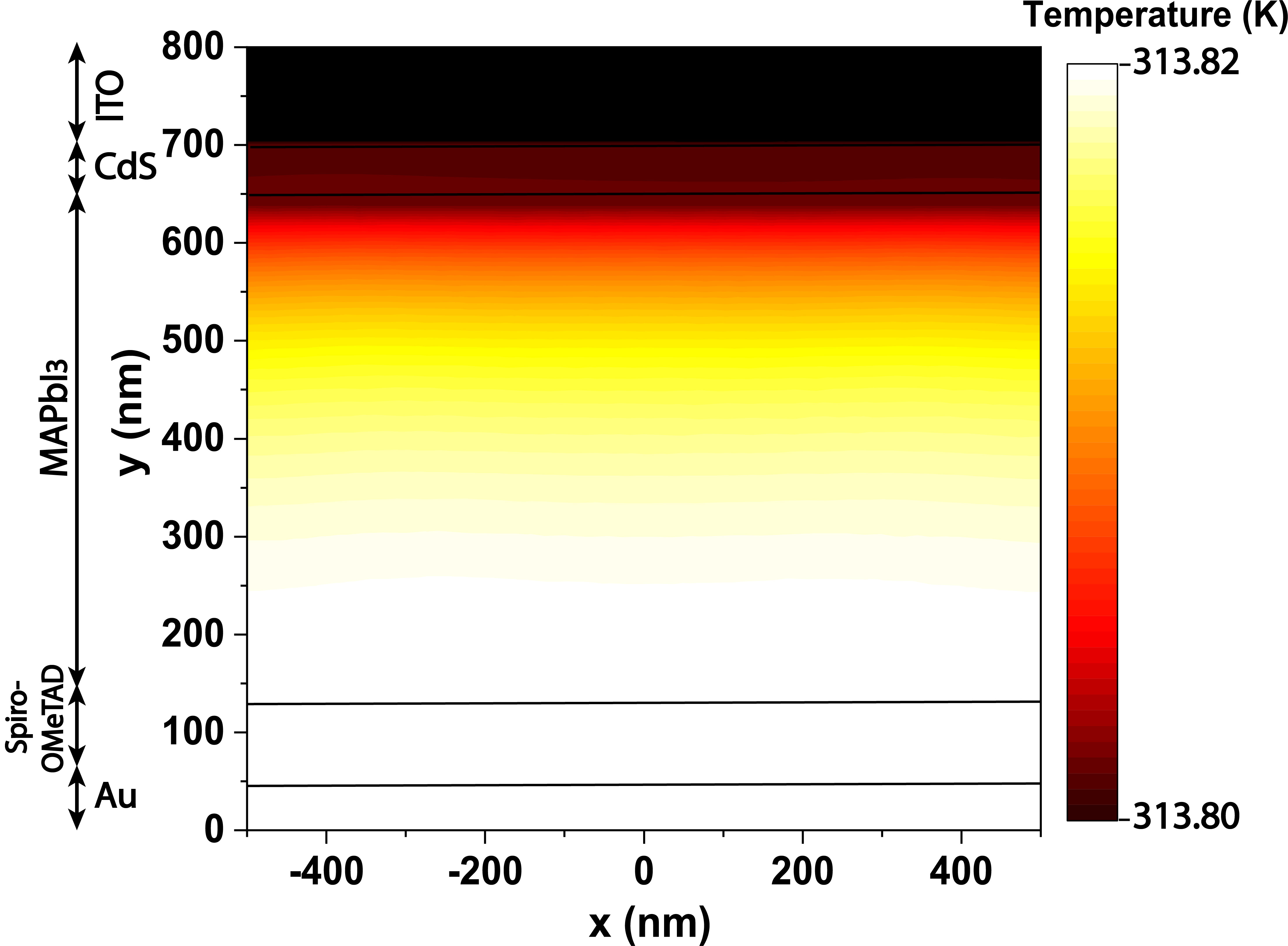}
        \caption{}
       \label{fig:temp_planer}
    \end{subfigure}
    ~
    \begin{subfigure}[b]{0.31\textwidth}
        \centering
        \includegraphics[width=\textwidth]{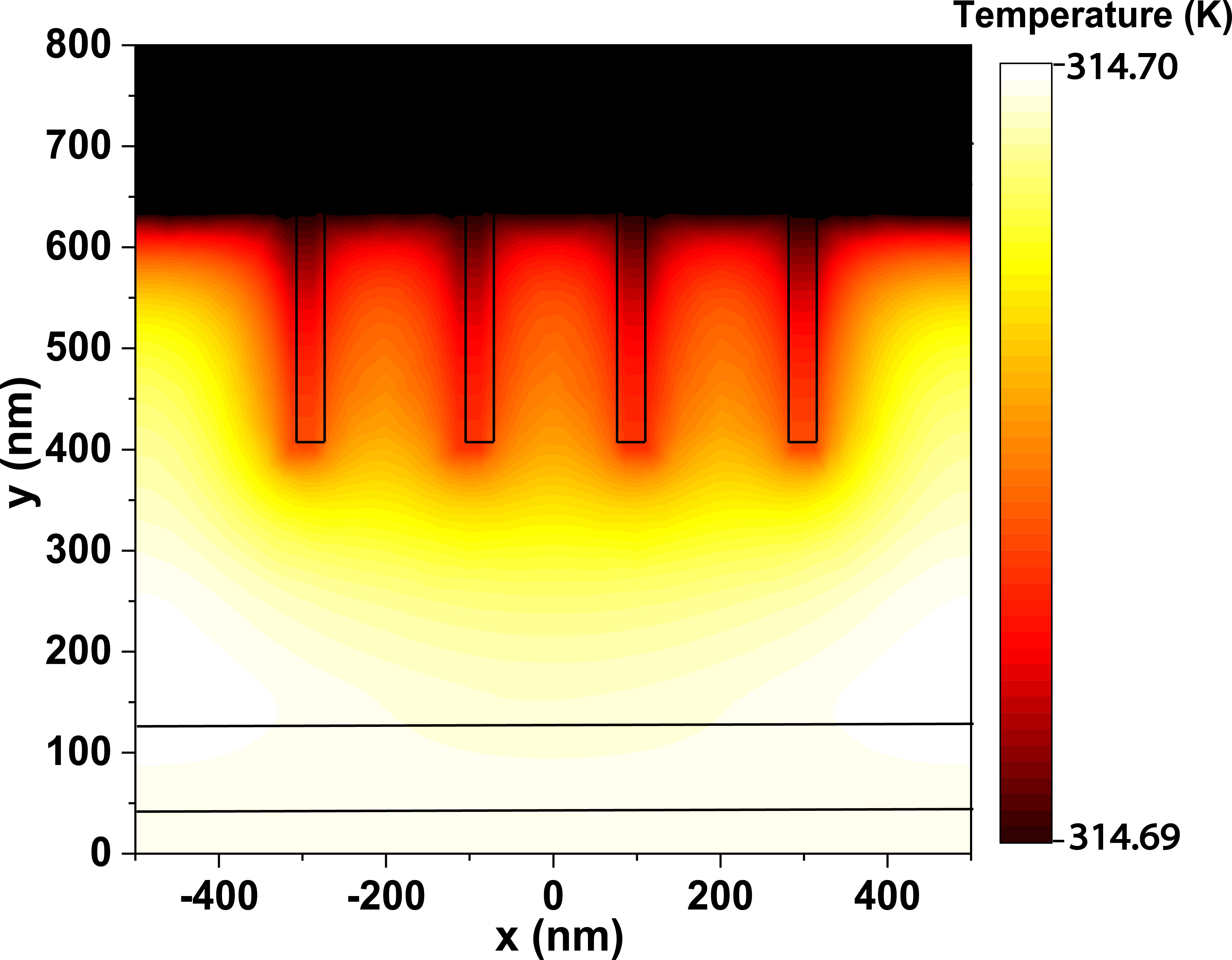}
        \caption{}
        \label{fig:temp_nanorod}
    \end{subfigure}
   ~
    \begin{subfigure}[b]{0.31\textwidth}
        \centering
        \includegraphics[width=\textwidth]{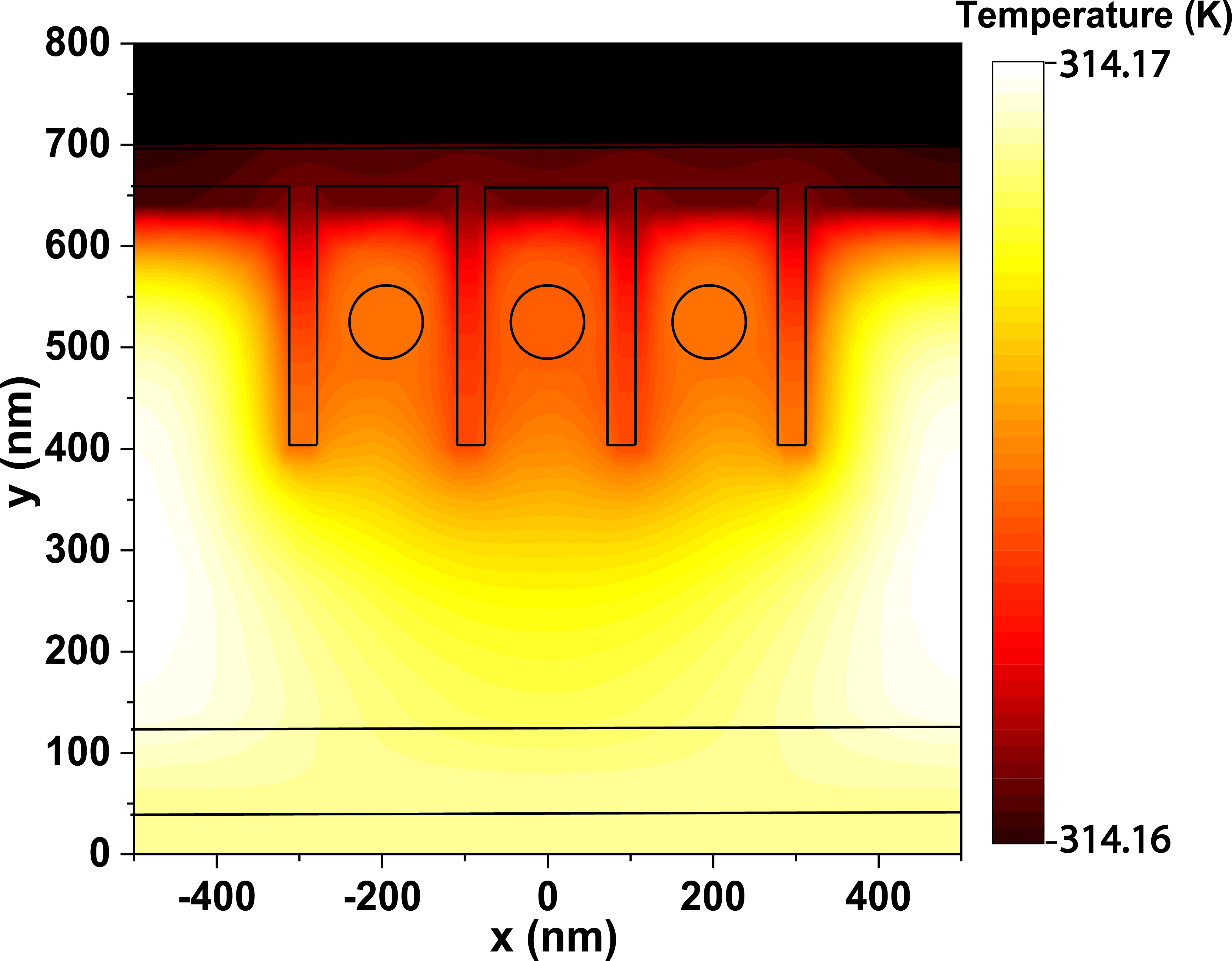}
        \caption{}
        \label{fig:temp_gold}
    \end{subfigure}
    
    \caption{ The temperature profiles of the (a) planar solar cell structure, (b) CdS grating structure, and (c) CdS grating structure with Au plasmonic nanoparticles have been investigated using thermal simulation. The temperature rise for all structures is approximately 14 K from the ambient temperature of 300 K.}
    \label{fig:temp_3struc}
\end{figure*}

All three structures, with an ambient temperature of 300 K, show a similar temperature rise of approximately 14 K. Figure \ref{fig:temp_planer} illustrates a temperature increase of 13.8 K for the planar structure, with a negligible temperature gradient from top to bottom due to the thin thickness of the solar cell. The temperature rise for the grated CdS and with Au NPs is approximately 14 K from the ambient temperature, as depicted in Fig. \ref{fig:temp_nanorod} and Fig. \ref{fig:temp_gold}, respectively. The CdS grating area in both of these structures exhibits a lower temperature compared to the surrounding perovskite. Figure \ref{fig:temp_gold} shows the thermal impact of adding the Au plasmonic NPs, altering the thermal profile near the Au NPs. The low rise in temperature indicates the reliability of the proposed device.

\subsection{Quantum Efficiency}
\begin{figure}[!ht]
    \centering
    \includegraphics[width=7.5cm]{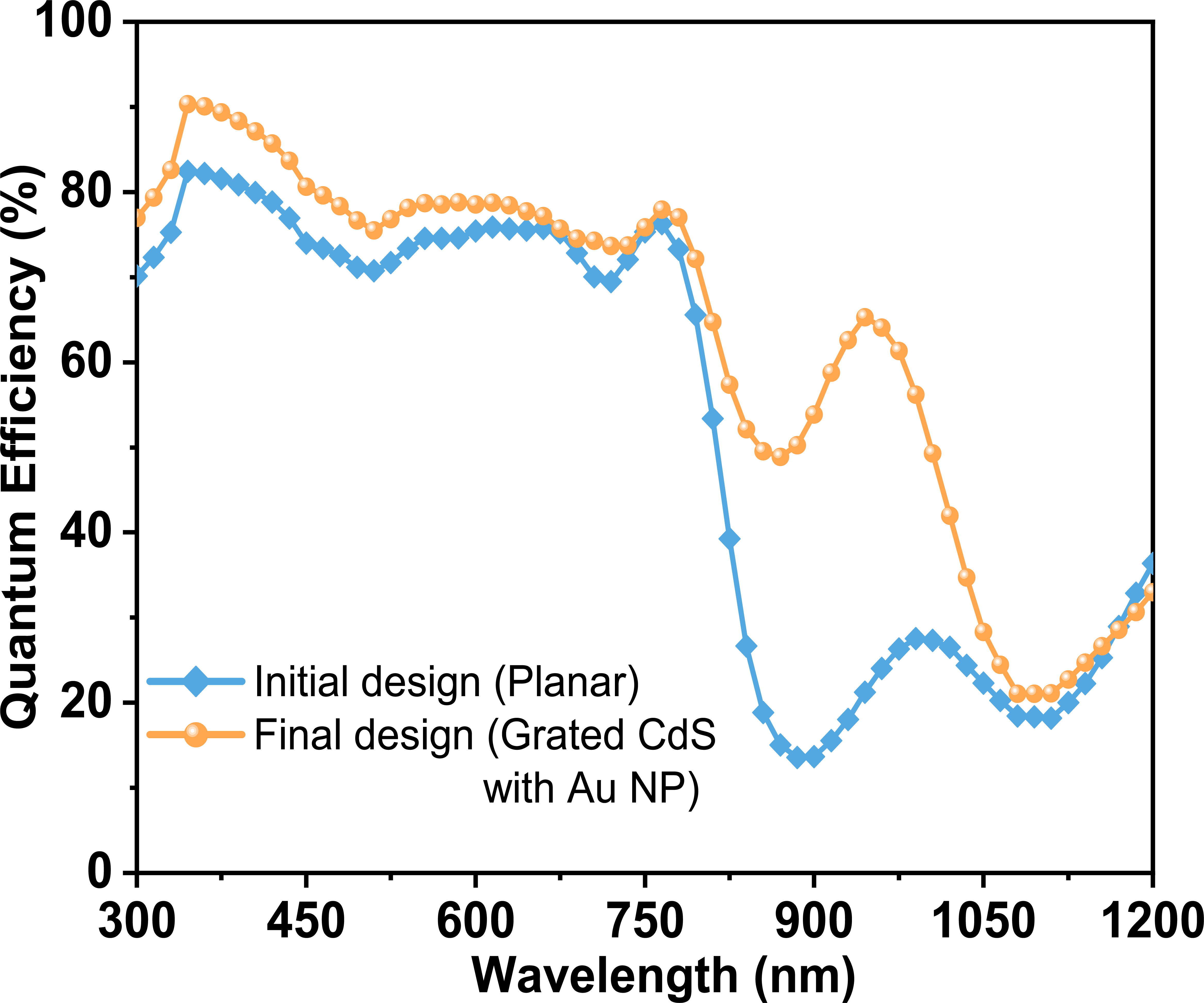}
    \caption{Extracted quantum efficiency (QE) spectra for the planar and grated CdS with Au NPs solar cell designs. The proposed grated CdS with Au NPs (red curve) structure exhibits significant QE improvement over a wide range of wavelengths (300 nm to 1200 nm) compared to the initial planar design (blue curve).}
    \label{fig:QE}
\end{figure}

The grated CdS with Au NPs solar cell design shows a significant improvement in quantum efficiency (QE) compared to the planar structure. In Fig. \ref{fig:QE}, the extracted QE spectra for the planar and grated CdS with Au NPs designs are shown. 

The grated CdS with Au NPs exhibits a notable improvement in QE over a wide range of wavelengths, from 300 nm to 1200 nm, compared to the planar structure. This improvement is due to the large contact surfaces between the CdS gratings and the perovskite material, which enhances carrier transport and reduces carrier recombination. Thus, the grated CdS with Au NPs shows an enhanced QE between 300 nm and 825 nm, where the planar structure exhibits poor QE. The incorporation of Au NPs in the grated CdS structure facilitates photon absorption and results in a significant improvement in QE beyond 780 nm. The proposed structure shows an improved QE between the wavelength range of 825 nm and 1020 nm since the QE is correlated with the absorbance and internal QE, and the grated CdS with Au NPs provides enhanced absorption beyond 825 nm \cite{abdelraouf2016towards}. At 945 nm wavelength, the QE for the grated CdS with Au NPs is around 65\%, while the QE for the planar structure at the same wavelength is approximately 21\%. Thus, the grated CdS with Au NPs design's combined impacts boost the QE, resulting in a substantial improvement in J\textsubscript{sc}.

\section{Conclusion}
In Conclusion, the proposed design in this study shows significant potential in addressing the challenge of low optical absorption rates in perovskite solar cells (PSCs), particularly in the infrared (IR) region. The initial model was a planar PSC device with an ITO/CdS/MAPbI\textsubscript{3}/Spiro-OmeTAD/Au structure. The incorporation of a grated CdS nanostructure significantly enhanced the overall optical absorption of the planar structure by granting an extended optical path, reducing electron-hole recombination rates, and increasing light absorption in the active layer. Furthermore, plasmonic gold (Au) nanoparticles were added between every two consecutive CdS grating plates in the absorber layer to improve optical absorption in the IR region (beyond 750 nm). The combined implementation of grated CdS and Au NPs structures resulted in a mean light absorption increase of 48\% in the 800-1400 nm range.

The electrical characteristics, including power conversion efficiency (PCE), fill factor (FF), open-circuit voltage (V\textsubscript{OC}), and short-circuit current (J\textsubscript{SC}), were evaluated alongside the optical absorption of the PSC. Moreover, the heat and temperature distribution throughout the solar cell were analyzed, and the impacts of the width of the CdS grated structure on the generation rate and the radius, shape, and position variation of gold NPs on device performance were investigated. The results showed that the incorporation of the nanostructures led to significant improvements of 24.2\% and 56.6\% for J\textsubscript{SC}, which increased to 7.42 mA/cm\textsuperscript{2}, and PCE, which increased to 7.91\%, respectively. The V\textsubscript{OC} value of the device also increased from 0.81V to 0.87V, and the FF increased by 9.33\%, from 56.25\% to 65.58\%, upon the implementation of the nanostructures. Therefore, our proposed device has the potential to pave the way for future research on incorporating nanostructures in PSCs to overcome the challenges associated with low optical absorption rates in the IR region, ultimately resulting in the development of more efficient solar cells.

\section{Funding}
No financial funding was granted for the research, authorship, or publication of this article.


\section{Acknowledgement}
The authors thank the technical support from Fab Lab DU and Semiconductor Technology Research Centre (STRC) at the University of Dhaka, and Bangladesh Research and Education Network (BdREN) for providing the necessary computational resources for this simulation study. In addition, the authors wish to express their gratitude to the Electrical and Electronic Engineering Department of Shahjalal University of Science and Technology(SUST) for their assistance.

\section{Declaration of Competing Interest}
The authors certify that they have no known financial or personal conflicts of interest.

\bibliography{bibliography.bib}

\end{document}